\documentclass[11pt]{article}
\usepackage[utf8]{inputenc}
\usepackage{graphicx}
\usepackage{subfig}
\usepackage{setspace}
\usepackage{mathtools}
\usepackage{amsmath}
\usepackage{nccmath}
\usepackage{pdfpages}
\usepackage{caption}
\usepackage{placeins}
\usepackage[margin=1in]{geometry}
\usepackage{hyperref}
\usepackage{natbib}
\usepackage{xcolor}
\usepackage{tikz}
\usetikzlibrary{shapes.geometric, arrows.meta, positioning}

\title{Dataset repurposing and disruptive AI research}
\author{
    Yulin Yu\textsuperscript{*1}, Yong-Yeol Ahn\textsuperscript{5}, Daniel M. Romero\textsuperscript{2,3,4} \\
    \textsuperscript{1} College of Information Science, University of Arizona, Tucson, AZ, USA \\
    \textsuperscript{2} School of Information, University of Michigan, Ann Arbor, MI, USA \\
    \textsuperscript{3} Center for the Study of Complex Systems, University of Michigan, Ann Arbor, MI, USA \\
    \textsuperscript{4} Computer Science and Engineering Division, University of Michigan, Ann Arbor, MI, USA \\
    \textsuperscript{5} School of Data Science, University of Virginia, Charlottesville, VA, USA \\
     \textsuperscript{*} Address correspondence to: yulinyu@arizona.edu\\
}

\begin{document}
\maketitle


\section*{Significance Statement}
Identifying novel uses of existing datasets can enable new scientific discoveries. Here, we find that dataset repurposing in the field of artificial intelligence is associated with more disruptive scientific work. However, most repurposed datasets are not widely adopted and receive less immediate attention. Meanwhile, the future success of data repurposing remains highly unpredictable. We also identify team characteristics associated with data-repurposing behavior. Overall, our findings suggest that data repurposing is a potential driver of scientific impact and provide insights for researchers, policymakers, and data curators.

\section*{Abstract}

Technological advancements are enabling increasingly systematic and large-scale data collection across all areas of science, driving scientific innovation.
In particular, AI research exemplifies this trend, having advanced rapidly through the assembly of massive datasets used to train and evaluate machine learning models.
However, the escalating demand for data, the difficulty of creating high-quality datasets, and the exhaustion of easily accessible data sources in AI research raise important questions about how to maximize the value of existing datasets through recombination and repurposing.
Here, we draw on two theoretical frameworks---\emph{recombinational novelty} and \emph{transformational creativity}---to examine the practice of data repurposing and its scientific impact.
Focusing on AI, we analyze scientific outcomes associated with data repurposing across more than 10,000 machine learning papers.
First, we find that although most repurposed datasets do not achieve broad visibility in the short term, data repurposing is associated with greater disruption.
Second, when repurposed data is adopted by subsequent research, the repurposing paper is associated with higher disruption and increased citation impact.
Third, repurposing teams tend to be more experienced, more institutionally prestigious, and involve academic--industry collaboration.
However, team characteristics poorly predict which repurposed datasets will be adopted by the community. These findings suggest that data repurposing may be an important approach to scientific discovery, and that its successful adoption is more common among larger teams and collaborations spanning academia and industry.


\section*{Introduction}

Modern scientific progress is increasingly powered by large, shared datasets, from the human genome and single-cell atlases in biology to sky surveys in astronomy and collider experiments in particle physics~\cite{yang2025standing}. Across these domains, the public release of scientific datasets has become a central engine of discovery, enabling researchers to build on common data resources and accelerate cumulative knowledge production~\cite{nagaraj2020improving,Gewin2016-cp,salganik2018bitbybit}. Beyond direct reuse, shared datasets can generate additional value when \emph{repurposed}---that is, used to address research questions different from those they were originally designed to answer. Such repurposing often extends datasets into contexts far removed from their initial scope. An example in the artificial intelligence field is ImageNet, originally developed for image classification benchmarking~\cite{deng2009imagenet}, which has since been repurposed for applications ranging from wildlife identification~\cite{Norouzzadeh2018} to medical imaging~\cite{shin2016deep} and, more recently, multimodal learning~\cite{radford2021learning}. This example raises a central question: does repurposing existing datasets systematically drive scientific innovation?

Despite growing recognition of the role of data in innovation, existing research has primarily established that data availability matters. Prior studies emphasize data sharing and reuse~\cite{pasquetto2017reuse,pasquetto2019uses,tenopir2011data,lafia2023direct,viswanathan-etal-2023-datafinder}, as well as efforts to measure the value of specific datasets within disciplines~\cite{nagaraj2022private,nagaraj2024mapping,nagaraj2024importance,nagaraj2023does,nagaraj2020improving}. Moreover, a growing body of work shows that open-access data can generate broader scientific and societal value, producing measurable impact across science, policy, and industry~\cite{colavizza2020citation,de2023influence}. Related studies also identify links between data use and scientific impact, showing, for example, that using public data can lead to more publications~\cite{nagaraj2020improving}.

However, these findings do not yet translate into actionable strategies for guiding researchers in how to use data more productively. A central unresolved question is how researchers creatively reuse existing data and whether different forms of creative data use are associated with different scientific outcomes. Although data recombination---the use of datasets that are not typically used together---has been identified as one pathway to higher scientific impact~\cite{yu2024does}, little is known about how broader forms of creative data use can be systematically measured or how they relate to innovation.

To address this gap, we develop measures to characterize different forms of creative data use and examine their associations with scientific impact. Characterizing these patterns provides an initial step toward understanding the role of data use in scientific innovation and generating hypotheses about the mechanisms underlying impactful research. In turn, these insights may help inform data contributors, curators, and publishers in developing practices and policies that better support high-impact scientific innovation.

Here, we focus on data repurposing as a mechanism through which existing datasets can generate new scientific value.
Repurposing---using something for a purpose it was not designed for---is a well-established driver of innovation in many domains.
For instance, aspirin was originally used to treat pain and fever, but was later repurposed for cardiovascular disease prevention~\cite{vane2003mechanism}.
As easily discoverable ``low-hanging'' compounds are exhausted, repurposing offers a way to extract additional value without the full costs of de novo development.
The same logic may apply to data: when datasets created for a specific purpose are repurposed to address research questions outside their original scope, does this catalyze the emergence of new research areas? 


Two theories of innovation and creativity illuminate how novel data use can drive scientific advances.
Recombinational theory posits that innovation often arises from recombining existing elements in unconventional yet intelligible ways~\cite{uzzi2013atypical, Foster2015-bm, lin2022new, leahey2023types, Shi2023-xv}.
Recent work shows that recombining datasets---typically within the same domain---is associated with higher scientific impact~\cite{yu2024does}.
However, prior studies have largely overlooked how data are applied across \emph{different} topics, leaving open the question of whether repurposing data for novel research questions contributes to innovation.
Transformational creativity theory offers a complementary lens: innovation can arise from altering the structure of a conceptual space, enabling fundamentally new ideas to emerge~\cite{boden1998creativity}.
This perspective suggests that datasets created for specific tasks (e.g., ImageNet for image classification~\cite{deng2009imagenet,gebru2021datasheets}) can be repurposed to open up entirely new research directions---yet empirical evidence for this remains limited.

Here, we study data repurposing in AI research, where data are the cornerstone for training and evaluating models and the exhaustion of easily accessible data sources makes creative reuse increasingly important. 
Consider ImageNet again: originally developed for image classification benchmarking~\cite{deng2009imagenet}, it has since been repurposed for transfer learning in medical imaging, wildlife identification from camera traps~\cite{Norouzzadeh2018}, and evaluating vision-language models---contributing to research areas far removed from its original intent. But is ImageNet's trajectory the exception or the rule?
We investigate whether data repurposing systematically yields high scientific impact by analyzing the relationship between the degree to which papers repurpose data and their outcomes.

We measure impact along two dimensions: disruptiveness, the extent to which a paper shifts the direction of its field, and citation count, a conventional proxy for scientific recognition.
Beyond the act of repurposing itself, we also ask whether repurposing reshapes how a dataset is used by subsequent research---that is, whether future papers adopt the repurposed application rather than the dataset's original use.
We refer to this phenomenon as \emph{repurpose propagation}. 
Finally, we examine what types of research teams engage in data repurposing and whether team characteristics predict this propagation.

We compiled a comprehensive dataset comprising more than 10,000 papers using 1,689 distinct datasets.
The primary data resource is Papers With Code (PWC)\footnote{\url{https://paperswithcode.com/datasets}}, an open-source repository of machine learning papers, datasets, and evaluation tables. 
PWC tracks which datasets are used in each study and has been used in prior research on AI data utilization~\cite{koch2021reduced,frank2019evolution,martinez2021research}.
The resulting corpus spans major AI venues including CVPR, ACL, and NeurIPS.
After sourcing these datasets from PWC, we linked the publication records to SciSciNet~\cite{lin2023sciscinet} and OpenAlex\footnote{\url{https://openalex.org/}}, which provide metadata and citation information for research papers.
A key advantage of PWC is that it records datasets that are substantively used in each study's analysis or model training, rather than merely cited---an important distinction given that data citations are often missing or do not reflect actual usage~\cite{fenner2019data,cousijn2019bringing,moss2018opaque,blake2010beyond}.
A complete description of our data is provided in the Materials and Methods section and SI Appendix, Section 1.


\section*{Quantifying data repurposing and scientific impact}

\begin{figure}[tbhp]
\centering
\includegraphics[width=.9\linewidth]{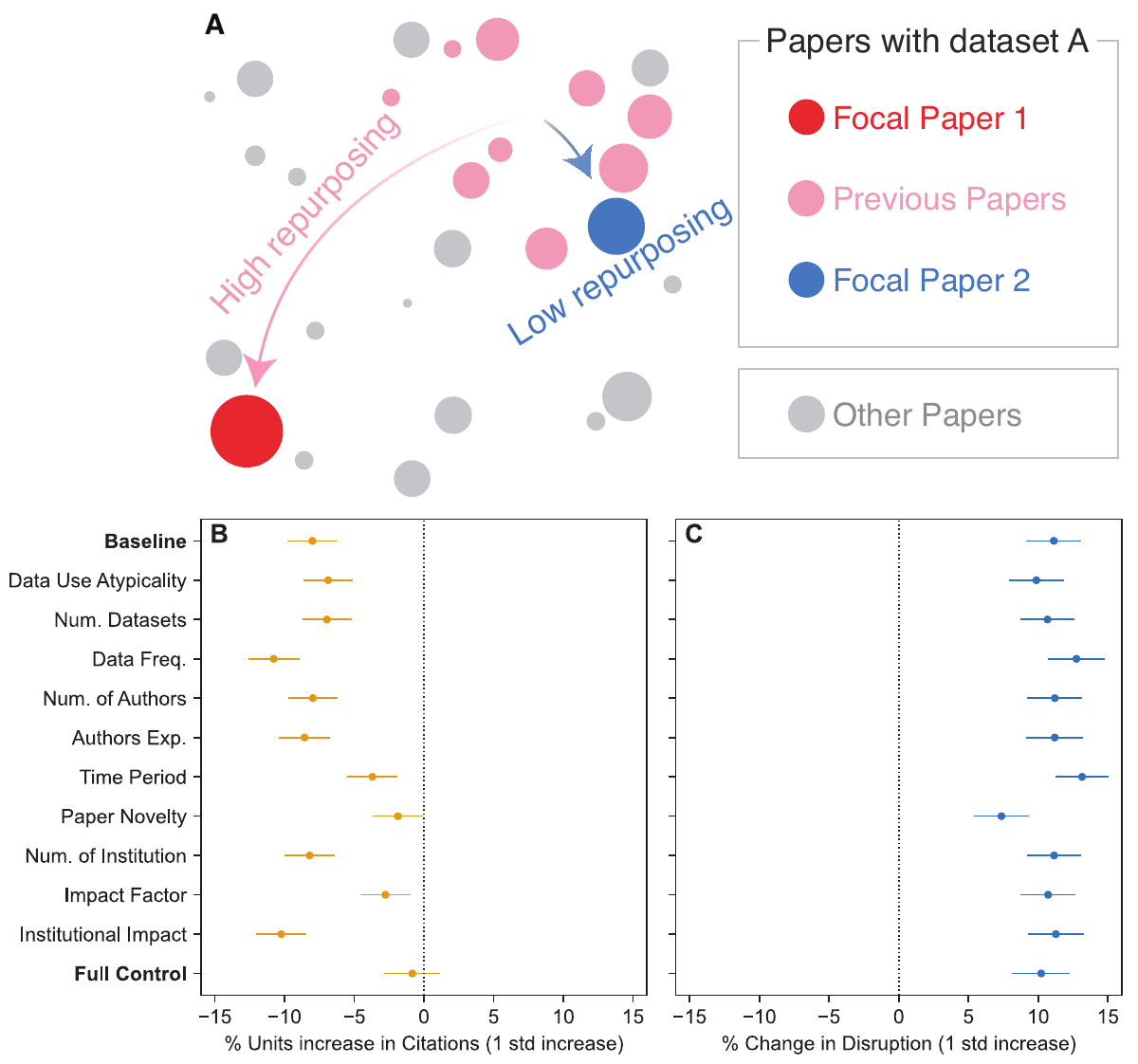}
\caption{
Data repurposing is associated with scientific disruption but not with short-term citation gain.
(A) Quantifying the degree of data repurposing. Each circle represents a paper positioned in a semantic embedding space, where distances reflect content similarity. Paper~1 shares dataset usage with papers shown in pink, while Paper~2 shares usage with papers in blue. Paper~1 has a higher repurposing score because it is more semantically distant from prior papers that used the same datasets.
(B) Effect size of \textbf{data repurposing} on \textbf{citation count} over 3 years, based on a negative binomial regression controlling for factors indicated in the panel headings. Each intermediate row reports the data repurposing coefficient from a separate
regression that includes only data repurposing and the control(s) named in that row. The baseline model includes no control variables; the full model includes all control variables combined. (C) Effect size of \textbf{data repurposing} on disruption score over 3 years, based on OLS regression with the same control structure as (B). 
Error bars represent 95\% confidence intervals.
} 
\label{fig:a1}
\end{figure}

To quantify data repurposing, we compare each paper's title with those of papers published at least one year earlier that used at least one of the same datasets.
To compare the semantic distance between two papers, we use embedding representations extracted from SPECTER2~\cite{Singh2022SciRepEvalAM} for each paper based on their titles and compute the distance between them.
The degree of data repurposing ranges from 0.009 to 0.659, with smaller values indicating less repurposing and larger values indicating greater data repurposing. Figure~\ref{fig:a1}(A) illustrates this measurement.
Further details on the operationalization, validation, and examples of this metric are provided in the ``Materials and Methods'' section and SI Appendix, Section 2.
To validate the measurement, we compared the data repurposing score with the judgments of two human annotators and found substantial alignment (73.1\% agreement; see ``Materials and Methods'').

We use two measures of scientific impact: disruption and citations.
To measure disruption, we use the $CD$ index~\cite{funk2017dynamic,Funk2012-bp,park2023papers,leahey2023types}, which captures whether a paper redirects or consolidates its field: if future works cite the focal paper but ignore its references, the paper is deemed disruptive; if they cite both, it is consolidating. For citation impact, we use raw citation counts.
Details on both measures are provided in the ``Materials and Methods'' section.

\section*{Papers that repurpose datasets tend to become more disruptive}

We use OLS regression to model the relationship between the disruption score and the degree of data repurposing.
We control for dataset-related variables: average data use frequency (as popular datasets may be linked to trending topics), the number of datasets, and the atypicality of data combinations~\cite{yu2024does}.
We also control for factors known to influence scientific outcomes: team size, team experience, publication year, and journal impact factor~\cite{wuchty2007increasing, mccain1990mapping, radicchi2008universality, redner1998popular}.
Following prior work linking atypical knowledge combinations to citation impact~\cite{uzzi2013atypical,Foster2015-bm}, we include the atypicality of references' journal combinations (``paper novelty'') as an additional control.
Since citation and reference counts are correlated with the disruption score, we also include these as controls~\cite{bentley2023disruption, petersen2025disruption, macher2024there,holst2024dataset}.
Full details are in the ``Materials and Methods'' section and SI Appendix, Section 3.

Our results show that repurposing datasets is associated with disruption but not with short-term citations. Papers with a higher degree of data repurposing achieve significantly higher disruption scores (Figure~\ref{fig:a1}(C); $p < 0.001$): A one standard deviation increase in data repurposing is associated with a 0.102 SD increase in the 3-year disruption score (VIF $=$ 1.176). This effect is robust to controls for paper novelty, team composition, journal characteristics, and dataset-related features. It is comparable in magnitude to the effect of dataset count and roughly half the magnitude of the effect of paper novelty. In contrast, a negative binomial regression with analogous controls finds no significant association between data repurposing and citation count (Figure~\ref{fig:a1}(B); slight negative point estimate, not significant in the fully-controlled model). Full regression tables are in SI Appendix, Section~4, Tables~1--2. The divergence is informative: Citations reflect attention and recognition, while disruption captures whether a paper shifts the norms and topics of its field, suggesting that data repurposing may not attract immediate attention but is associated with subsequent shifts in research direction.


\section*{Repurpose propagation is less likely but more consequential}

When a paper takes a dataset in a new topical direction, subsequent work may either follow this new direction or remain close to earlier uses. In the former case, the repurposing paper may shift how the research community perceives the dataset’s applicability; alternatively, the original use cases may be exhausted, such that the shift reflects broader field dynamics rather than the paper itself.  In this section, we quantify the extent to which a repurposing paper coincides with subsequent use of the dataset in the new topic area. We call this phenomenon \emph{repurpose propagation}.

\begin{figure*}[tbhp]
\centering
\includegraphics[width=.99\linewidth]{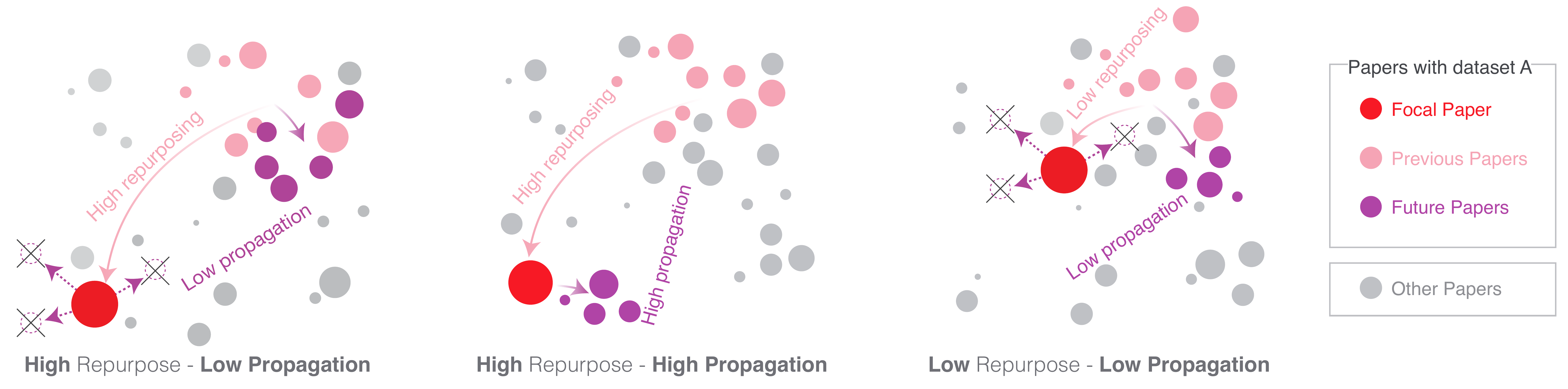}
\caption{Illustration of repurpose propagation in a 2D semantic space. Circles in pink denote papers published before the focal paper (red circle); circles in purple denote papers published after the focal paper using the same dataset. In the {\bf{high-repurpose, low-propagation case}}, the focal paper is semantically distant from both prior and subsequent papers. In the {\bf{low-repurpose, low-propagation case}}, the focal, prior, and subsequent papers cluster in the same region. In the {\bf{high-repurpose, high-propagation}} case, the focal paper is far from prior papers but close to subsequent papers. A positive result indicates that the data was subsequently repurposed in future papers; a negative result suggests otherwise.
}
\label{fig:a33}
\end{figure*}

We define repurpose propagation as the difference between a focal paper’s weighted topical similarity to future papers using the same dataset and its weighted topical similarity to prior papers using that dataset. Positive values indicate that future papers move toward the topical direction introduced by the focal paper, whereas negative values indicate that future papers remain closer to prior uses of the dataset. Because this measure is the difference between two cosine similarities, its theoretical range is $[-2,2]$, though empirically the values are concentrated within a much narrower range. Figure~\ref{fig:a33} illustrates this measurement and shows examples of different combinations of repurposing and propagation; see Materials and Methods for details.

\begin{figure}[tbhp]
\centering
\includegraphics[width=.9\linewidth]{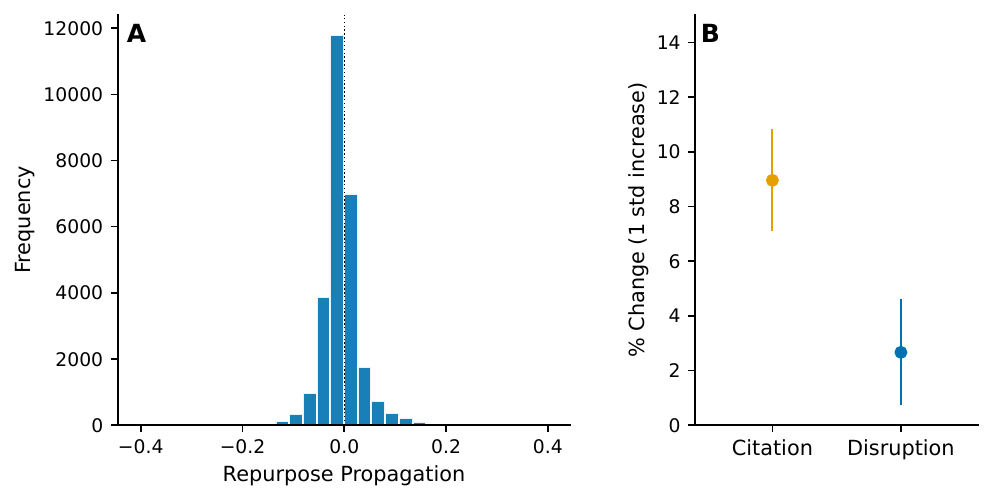}
\caption{Although repurposed data tends not to be propagated, propagation instances yield substantial impact.
(A) Distribution of repurpose propagation. (B) This plot illustrates the effect size of the repurpose propagation variable on the disruption score over three years, based on an OLS regression controlling for all factors (blue), and on the citation score over the same period, based on a negative binomial regression also controlling for all factors (orange).}
\label{fig:a3}
\end{figure}

Figure~\ref{fig:a3}(A) shows that data repurposing propagates only weakly on average, as the distribution of repurpose propagation scores is shifted slightly below zero. Specifically, 55\% of papers exhibit a repurpose propagation score smaller than 0 ($p < 0.001$, one-proportion sign test against $\Pr(A_a < 0)=0.5$). The median repurpose propagation score is $-0.008$, with an interquartile range of $[-0.023, 0.008]$.

We then follow the same modeling approach used for examining the relationship between data repurposing score and both disruption score and citation count, using an OLS model for disruption score and a negative binomial model for citations.
We observed that papers with a higher rate of repurpose propagation also tend to be more disruptive. Specifically, a one-standard-deviation increase in repurpose propagation is associated with a 0.03-standard-deviation increase in the disruption score ($p < 0.001$), although the effect size is smaller than that of the repurposing score alone.
However, it is associated with a significant increase in future citations.
A one standard deviation increase in repurpose propagation score is associated with an approximately 9\% increase in citations ($p < 0.001$).
Figure~\ref{fig:a3}(B) displays the regression coefficients concerning the relationship between repurpose propagation and both disruption and citation, with full control variables (the complete regression table is available in the Supplemental Information (SI) Appendix, Section 4, Tables 3--4).

\section*{Team characteristics associated with repurposing, but not with propagation
}

Given that data repurposing is associated with disruptive scientific work, our final analysis aims to identify which types of research teams are more likely to repurpose data and do so with future propagation.
Prior studies have highlighted the critical role of teams and their composition in fostering scientific innovation and have particularly emphasized the impact of team size~\cite{uzzi2013atypical,wu2019large}, academic-industry collaboration (especially in AI research)~\cite{ahmed2023growing}, and global collaboration~\cite{graves2022inequality}.
Additionally, the age or experience of authors is often associated with creativity and innovation~\cite{jones2009burden}.
However, how team composition shapes repurposing and its propagation remain an open question.
We use Ordinary Least Squares (OLS) regressions to model the relationships between the data repurposing score and various team composition variables in separate models: 1) team size, 2) team experience, measured by the authors’ average citation
count, 3) institutional prestige in research measured by the average citation count of the paper's affiliated institutions, 4) global status of the institution's country (categorized as developed, developing, or mixed collaboration following the United Nations classification of economies\footnote{United Nations Department of Economic and Social Affairs, \textit{World Economic Situation and Prospects}. \url{https://www.un.org/development/desa/dpad/publication/world-economic-situation-and-prospects/}}), and 5) the team's sectoral background, including industry, academia, or a combination of both, determined by the authors' affiliations.
We also model repurpose propagation score against these five team characteristics.

\begin{figure}
\centering
\includegraphics[width=.9\linewidth]{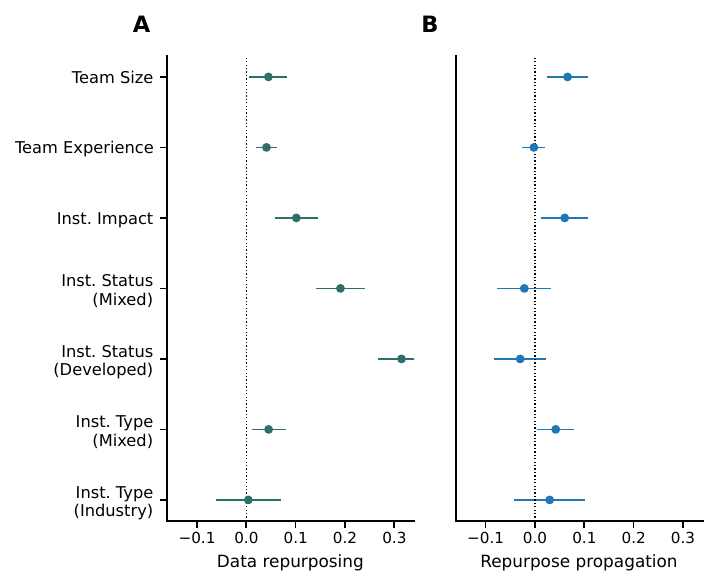}
\caption{Team characteristics are associated with data repurposing more than with its propagation. (A) presents ordinary least squares (OLS) regression results examining the relationships between team size, team experience, institutional impact, institutional global status, and institutional type (industry vs.\ academia) and the degree of data repurposing. (B) examines the relationships between the same variables and the degree of repurpose propagation. We estimate a separate regression model for each listed variable, while controlling for data recombination, the number of datasets used, and dataset usage frequency. Markers represent regression coefficient estimates, and error bars indicate 95\% confidence intervals.}
\label{fig:a4}
\end{figure}

We find that team characteristics are much more strongly associated with the likelihood of repurposing data, whereas they show little to no significant association with successfully conducting data repurposing that is later adopted by the research community. Figure~\ref{fig:a4}(A) shows the regression coefficients modeling the relationship between data repurposing score and various team composition features, and Figure~\ref{fig:a4}(B) shows the coefficients of repurpose propagation with team composition features. Six of the seven team-related coefficients are significantly associated with creating papers that repurpose data. The seven team-related measures include team size, team experience, institutional impact, whether institutions are from industry or academia, mixed industry--academia collaborations, developed versus developing countries, and mixed developed--developing country collaborations. In contrast, only three team-related variables are significantly associated with successfully repurposed data, and their effect sizes are substantially smaller.Full regression tables are
in SI Appendix, Section 4, Tables 5–212.

Several findings stand out. First, while more experienced teams are associated with producing papers that repurpose data, their experience is not significantly associated with repurpose propagation. Second, team size is significantly associated with both data repurposing and whether repurposed data are adopted. Second, researchers from developed countries are more likely to produce repurposing papers, but country development status is not significantly associated with propagation. In other words, the development status of a researcher's country is not significantly associated with the success of data repurposing. Fourth, notably, both academic--industry collaboration and institutional research impact are significantly associated with data repurposing and its propagation.



\section*{Discussion}
Motivated by the rapid advancement of AI and its reliance on publicly available data, we analyze dataset usage across more than 10,000 AI papers spanning 1,689 datasets and report three main findings. First, we find that repurposing datasets is associated with more disruptive papers, although such papers receive limited short-term citation attention. Second, we find that repurpose propagation is associated with greater disruption and higher citation impact. Third, team characteristics are strongly associated with whether data repurposing occurs, but most do not predict its subsequent propagation. Notable exceptions are industry--academia collaboration, institutional research impact, and team size, all of which are strongly associated with repurpose propagation. In particular, the positive association between team size and propagation provides an interesting counterpoint to prior work showing that smaller teams are more likely to produce disruptive research~\cite{wu2019large}: while smaller teams may be better positioned to pursue disruptive ideas, larger teams may have greater capacity to facilitate the subsequent diffusion and adoption of repurposed datasets.

Our results speak to two theories of novelty. First, they align with prior research showing that knowledge recombination is associated with disruptive outcomes~\cite{lin2022new}. In addition, novelty in methods, theories, and results has also been linked to disruption~\cite{leahey2023types}. We contribute to the science of science literature by showing that recombining distinct knowledge modalities---particularly topics and data---is associated with disruptive scientific contributions. Second, we provide empirical support for transformational creativity theory by showing that data repurposing is associated with disruptive scientific contributions.
    
The absence of short-term citation gains resembles ``pivot'' behavior, which has been linked to a short-term novelty penalty~\cite{hill2025pivot}. However, prior work on pivoting focuses on \emph{authors} changing their own research direction, where the penalty arises from entering a new domain. In contrast, data repurposing does not necessarily imply an author-level pivot: authors may either move into a new topic using an existing dataset (where a pivot penalty may apply) or remain within their domain while adopting data from another field (where no such penalty would be expected). As such, it remains unclear whether the pivot-based mechanism fully explains the absence of short-term citation gains. Future work could disentangle these pathways by comparing cases where authors shift topics versus those where datasets are repurposed across domains without author-level pivots.

Even high-repurposing papers typically do not propagate. This finding resonates with prior research suggesting that data can narrow scientists’ imagination, leading them to prioritize what is readily accessible over exploring higher-value but less obvious directions---a tendency known as the streetlight effect~\cite{Hoelzemann2024-ia}. A complementary explanation is that broader propagation may only occur once a dataset’s value within its original domain is exhausted, prompting researchers to seek applications in new contexts. Because such exploration is difficult and uncertain, it is likely to be relatively rare. Together, these potential reasons shape how data are paired with topics: researchers may default to a dataset’s established uses or rely on familiar dataset--topic pairings, thereby limiting the extent to which repurposing propagates.

Our results also speak to the relationship between team size, disruption, and data use. Prior work has shown that small teams are more likely to produce disruptive research~\cite{wu2019large}, and other studies have found that smaller and less experienced teams are more likely to repurpose data in novel ways, particularly through unusual dataset combinations in data science research~\cite{yu2024does}. Our study adds to this literature from the perspective of data use: we conceptualize repurpose propagation as a disruptive way of using data, measured by whether future papers using the same dataset are semantically closer to the focal paper than to prior papers using that dataset. In contrast to prior work emphasizing team effects, we find that while team characteristics are associated with the act of data repurposing itself, most do not predict repurpose propagation. The primary exceptions are industry--academia collaboration and institutional research impact, both of which are positively associated with repurpose propagation. One possible explanation is disciplinary differences. Innovations stemming from data discovery in fields such as the social sciences often require extensive reflection and iteration between research questions and datasets, as well as substantial time to uncover meaningful insights. By contrast, AI innovation often relies more heavily on collaboration and computational resources to explore new directions. Together, these findings suggest that while team structure may shape the likelihood of attempting novel data use, the broader propagation of repurposed ideas cannot be fully explained by team characteristics alone.

Our study offers several implications.
First, if the association reflects a causal effect, it may be possible to build an environment that nudges researchers to consider repurposing datasets. With advancements in large language models, future tools could actively support data repurposing by identifying promising dataset--topic combinations based on factors highlighted in our study, such as topic distance and prior dataset usage patterns. For example, dataset repositories could surface papers from distant research areas that previously reused similar datasets, conference submission systems could flag datasets that have only been applied within narrow topical domains, and LLM-based recommendation systems could suggest unexplored but potentially fruitful dataset applications across fields. Second, data curation practices, publication policies, and repositories could include summaries of the research topics in which datasets have been used and introduce prompts or recommendations when a dataset is repeatedly applied within a narrow scope, as such patterns may limit the disruptive potential of subsequent research~\cite{Hill2020-wo,Heberling2021-ge}.

Our study has five main limitations. First, Papers With Code, while an excellent resource for computer science, lacks clear inclusion criteria and is concentrated in mainstream venues such as CVPR and ACL, which may limit representativeness across the broader AI research landscape. Second, constructing the analytical sample required excluding papers with inaccessible metadata, potentially introducing selection biases related to factors such as country, institution, or publication venue. Third, our study examines only short-term disruption and citation outcomes over a three-year window due to the recency of the Papers With Code dataset and limited data availability before 2015. Because computer science evolves rapidly, longer-term follow-up studies are needed to distinguish transient from durable effects~\cite{wang2013citation}. Fourth, the disruption measure may be mechanically related to our data repurposing measure. Repurposing papers are often topically distant from earlier work using the same dataset, so they and their subsequent citers may be less likely to cite those earlier papers, mechanically increasing the disruption index. However, this mechanical account does not fully explain our findings. Repurposing papers almost always cite the original source of the dataset, meaning that subsequent papers could have continued citing both the original work and the repurposing paper, resulting in a bridging rather than redirecting citation pattern. The observed tendency toward redirecting citation patterns therefore reflects how the research community subsequently builds on repurposed data, rather than a mechanical consequence of our measure. Moreover, our repurposing measure is computed solely from paper content before any downstream citation behavior exists, whereas the disruption index is determined by citation patterns that emerge only afterward. The association between data repurposing and disruption therefore reflects an empirical relationship between data-use behavior and subsequent scientific influence, rather than a bookkeeping artifact. That said, although the association remains robust after controlling for reference and citation counts, these controls primarily address known artifacts of the disruption index and do not completely eliminate the possibility of mechanical overlap. This concern may be particularly relevant given our three-year citation window. If repurposing introduces a dataset into a new topical community, early citations may disproportionately come from that new community and cite the repurposing paper without yet connecting back to the broader dataset lineage. Such cross-community bridging may take longer to emerge, potentially producing a relatively high short-term disruption score even when repurposing ultimately facilitates integration between previously disconnected research communities. Because the disruption index can be sensitive to the citation window over which it is measured, our results should therefore be interpreted as evidence of short-term disruptive influence rather than necessarily persistent disruption. Future work using longer citation windows could examine whether the association between data repurposing and disruption persists or attenuates as these bridging citation patterns have more time to develop. Fifth, the observed association between propagation and citation impact may also contain a mechanical component. Our propagation measure is computed from future papers that reuse the same dataset and are semantically similar to the focal paper, and these papers are also among those most likely to cite it. Given the highly skewed distribution of citations, even a small number of such highly probable citers could meaningfully contribute to citation counts. Future work could further evaluate this possibility by excluding papers in the propagation set from citation counts or by measuring citation impact over a later time window that does not overlap with the propagation window.

For future research, we recommend further experiments or causal studies to strengthen the evidence regarding data repurposing. It would also be valuable to investigate why data repurposing remains relatively uncommon and why team characteristics predict the act of repurposing but not the later propagation of repurposed datasets. Additionally, given substantial disciplinary differences in data curation, usage, and citation practices, future work should examine whether the patterns identified here generalize across fields~\cite{Tenopir2020-um,gregory2023tracing,jiao2023exclusively}. Although not all instances of data repurposing will gain widespread adoption, our findings suggest that repurposing can open research directions that depart from established patterns of dataset use and contribute to scientific innovation. Our findings further suggest a hypothesis for future testing: interventions that lower the barriers to initiating data repurposing—for example, through improved dataset discovery, documentation, interoperability, and training in cross-domain data use—may be particularly valuable, given that team characteristics are more strongly associated with the initiation of data repurposing than with its subsequent propagation.


\section*{Materials and Methods}

\subsection*{Data}

Our final analysis uses a dataset of 12,286 artificial intelligence papers published in or before 2021, derived from Papers With Code (PWC)~\cite{koch2021reduced}. The majority of paper records are from 2015 to 2021. The PWC dataset primarily indexes benchmark-driven research, where datasets are used to evaluate computational model performance. As such, it likely overrepresents benchmark-oriented, performance-focused studies and underrepresents other forms of AI research, such as theoretical or non-benchmark empirical work. This sampling bias should be considered when interpreting our results.

To construct the analytical sample, we first link PWC papers to OpenAlex\footnote{\url{https://openalex.org/}}
 in order to obtain publication metadata---including citation counts, discipline, publication year, impact factor, references, team size, and author experience (measured by authors’ average prior citations). The initial PWC dataset contains 60,647 publications, of which 20,016 could not be matched to OpenAlex. These unmatched papers are disproportionately older papers and publications from non-traditional venues, potentially biasing the final sample toward more recent and well-indexed work. Among the remaining 40,631 papers, 35,801 published in or before 2021 contain the concept tags and reference data required to compute impact measures and control variables.

We further restrict the sample to published conference and journal papers, excluding 19,850 papers without venue information, the vast majority of which are arXiv preprints. Robustness checks including arXiv papers produce qualitatively similar results. Finally, we restrict the sample to papers with the opportunity for dataset reuse, defined as papers for which at least one dataset had been used in a publication more than one year earlier, and for which at least one publication more than one year later also reused a shared dataset. This step removes 3,665 papers that lack either prior or future dataset usage. The final analytical sample consists of 12,286 conference and journal publications. We use 2021 as the cutoff year to ensure at least three years of citation data for impact measurement. Additional details are provided in SI Appendix, Section 1 and SI Figure 5 (“Data construction flow chart for the analytical sample”).

\subsection*{Measuring dataset repurposing}

Let $\mathcal{D} = \{ d_1, d_2, \dots, d_n \}$ denote the universe of all datasets.
Each paper $a$ uses a subset of datasets $D_a \subseteq \mathcal{D}$. We define \emph{dataset repurposing} as a continuous measure rather than a binary indicator.
For a focal paper $a$ published in $y_a$, consider $P_a$, the set of previously published papers that share at least one dataset with paper $a$:
\[
P_a = \{\, b \mid D_a \cap D_b \neq \emptyset \;\text{and}\; y_b \leq y_a - 1 \,\}.
\]

To quantify the degree of repurposing, we measure how much the knowledge content of paper $a$ diverges from that of papers in $P_a$, while accounting for partial overlap in dataset usage.

First, to quantify the ``distance'' from prior works, we represent each paper as an embedding vector in a semantic knowledge space constructed using SPECTER2~\cite{Singh2022SciRepEvalAM}, a widely-used scientific document embedding model that captures document-level semantics from citation and textual information~\cite{cohan2020specter}.
We then quantify the similarity between two papers, $S_{ab}$, using the cosine similarity between the embedding vectors of the titles of papers $a$ and $b$.
While the SPECTER2 embedding is specifically trained on academic papers, its use may implicitly incorporate future citation information. Specifically, SPECTER2 is trained using citation relationships, such that papers sharing future citations are positioned closer in embedding space. This creates two related concerns in our setting. First, when computing the repurposing score $R_a$, the focal paper $a$ may appear artificially closer to prior papers $b \in P_a$ if later papers cite both, thereby lowering the estimated repurposing score for papers whose repurposing was ultimately influential. Second, when computing propagation, the focal paper may appear artificially closer to future papers through the same citation-driven mechanism, inflating estimated propagation. In both cases, the embedding representation partially encodes the downstream citation patterns that our measures aim to evaluate. To address this concern, we adopt an alternative approach by generating sentence embeddings from paper titles using the OpenAI embedding model (text-embedding-3-small), which is trained to capture general semantic relationships rather than citation-network structure and is therefore less likely to directly encode downstream citation patterns. The results remain qualitatively similar, with all estimates pointing in the same direction as those from our primary analysis (see SI Note 5).

Because prior papers may use only a subset of the datasets employed by the focal paper, we weight the semantic similarity by the extent of dataset overlap.

Specifically, we use the Jaccard similarity between the set of datasets in paper $a$ and that of $b$ to quantify the dataset overlap as follows:
\[
J_{ab} = \frac{|D_a \cap D_b|}{|D_a \cup D_b|}.
\]

Now, the degree of dataset repurposing for paper $a$, denoted $R_a$, is defined as a weighted divergence from prior work, where semantic similarity is weighted by dataset overlap, as formalized in Equation~\ref{eq:novelty1}.
\begin{equation}
  R_a= 1- \frac{\sum_{i \in P_a} S_{ai} J_{ai}}{\sum_{i \in P_a}  J_{ai}}
  = \frac{\sum_{i \in P_a} (1 - S_{ai})\, J_{ai}}{\sum_{i \in P_a}  J_{ai}}.
\label{eq:novelty1}
\end{equation}
 Since $S_{ai} \in [-1, 1]$, $R_a$ is theoretically in $[0, 2]$, with $0$ indicating no repurposing and $2$ indicating the highest level. However, in practice, the repurposing score lies in $[0, 1]$, since SPECTER2 similarities for scientific papers are usually positive.
Intuitively, $R_a$ is larger when a paper uses shared datasets to explore more novel regions of the knowledge space, particularly when overlap with prior work is limited.
For regression analyses, we standardize $R_a$ to account for scale differences and to improve comparability across variables.

\subsection*{Repurpose propagation measurement}

We next define \emph{repurpose propagation}, which measures how much a focal paper $a$ shifts the subsequent usage of its datasets relative to prior usage.
Like $P_a$ as defined above, let
\[
F_a = \{\, b \mid D_a \cap D_b \neq \emptyset \;\text{and}\; y_b \geq y_a + 1 \,\}
\]
denote the set of \emph{future} papers that share at least one dataset with paper $a$.

Using $S_{ab}$ and $J_{ab}$ as defined above, we compute the weighted semantic similarity between the focal paper and prior papers as
\[
\bar{S}^{\text{prior}}_a = \frac{\sum_{b \in P_a} J_{ab} \cdot S_{ab}}{\sum_{b \in P_a} J_{ab}},
\]
and analogously, the weighted semantic similarity between the focal paper and subsequent papers as
\[
\bar{S}^{\text{future}}_a = \frac{\sum_{b \in F_a} J_{ab} \cdot S_{ab}}{\sum_{b \in F_a} J_{ab}}.
\]

We define the degree of repurpose propagation for paper $a$ as 
\begin{align}
A_a &= \bar{S}^{\text{future}}_a - \bar{S}^{\text{prior}}_a \nonumber \\
    &= \frac{\sum_{j \in F_a} J_{aj} \cdot S_{aj}}{\sum_{j \in F_a} J_{aj}}
     - \frac{\sum_{i \in P_a} J_{ai} \cdot S_{ai}}{\sum_{i \in P_a} J_{ai}}.
\label{eq:novelty2}
\end{align}
Intuitively, if the focal paper $a$ repurposes datasets in a novel content area and this repurposed usage is adopted by later work, then future papers using the same datasets should be more semantically similar to $a$ than earlier papers. However, this pattern is observational and does not imply that $a$ caused the shift, as high $A_a$ may also reflect broader field-level movement toward the same topic. 
A larger value of $A_a$ indicates that subsequent papers using the same datasets align more closely with the focal paper's content than prior work, suggesting that paper $a$ has influenced how the datasets are applied across research topics. Focal papers with no future dataset reuses ($|F_a| = 0$) are excluded from the calculation of $A_a$. 

\subsection*{Validating repurposing measurement}

We validate the data repurposing score by testing whether the scores align with human judgments.
In this evaluation, two participants---namely one author and one master's student with sufficient research training---were presented with 2--4 papers and asked to determine which paper repurposed the shared dataset more extensively relative to prior papers using the same dataset. To make this evaluation feasible, we first select datasets that have been utilized in three to five papers, since the task requires evaluators to read all paper titles associated with a given dataset.
Within each set of papers corresponding to a dataset, we compute the variance of their repurposing scores and select the top 30\% of datasets with the highest within-year score variance, ensuring that the selected comparisons are discernible by evaluators. We note, however, that this procedure intentionally selects clearer high-contrast cases by comparing the highest- and lowest-scoring papers within high-variance subsets. As a result, the reported evaluation accuracy should be interpreted as performance on relatively distinguishable cases and may overestimate accuracy for more typical or ambiguous comparisons.
For each dataset in the top 30\%, we retain at least two papers published in the same year with the highest and lowest repurposing scores.
We also include earlier papers that used the same datasets as context for this exercise. This procedure yields 39 groups of papers for evaluation.

Two evaluators independently evaluate each set of papers (grouped by dataset).
For each dataset, they review the titles of papers published prior to the target papers and determine which paper demonstrates a higher degree of data repurposing.
From these evaluations, we report: (1) the level of agreement between the two evaluators, (2) the level of agreement between the evaluators and the outcome inferred by the computed repurposing score (noting that ``unable to determine'' counts as disagreement), and (3) the level of agreement considering only cases where both evaluators reached the same judgment.
Our results show that, on average, agreement between the evaluators is 71.2\%; the computed repurposing score aligns with human judgments in 73.1\% of all comparisons and in 82.1\% of comparisons on which both evaluators agreed.


\subsection*{Measuring disruption of publication}

The disruption index measures the impact a paper has on modifying the scholarly landscape~\cite{wu2019large,park2023papers,funk2017dynamic}.
This index, denoted $CD$ index, is derived from analyzing citation networks up to 3 years after publication.
For each focal paper, subsequent papers are categorized into three groups: those that cite only the focal paper, those that cite both the focal paper and its references, and those that cite the references without citing the focal paper.
The formula for $CD$ index is defined as follows:
\begin{equation}
CD = \frac{n_{i} - n_{j}}{n_{i} + n_{j} + n_{k}},
\label{eq:novelty3}
 \end{equation}
where $n_{i}$ is the count of later publications that cite only the paper in question,
$n_{j}$ is the count of those that cite both the paper and its references, and
$n_{k}$ is the count of those that cite only the references of the paper.
Using this methodology, we follow~\cite{lin2023sciscinet} and compute disruption scores for papers that have received at least one citation and contain at least one reference. $CD$ index has been criticized for being sensitive to the number of references and citations~\cite{bentley2023disruption,petersen2025disruption,macher2024there,holst2024dataset}. Therefore, we include the number of references and citations as controls when examining the relationship between data repurposing, repurpose propagation, and disruption in order to ensure the robustness of our results. The CD index is standardized in our estimation.

\section*{Acknowledgements}
We are deeply grateful for the help and suggestions provided by Misha Teplitskiy, Yian Yin, Kentaro Toyama, members of the Romero group, and the seminar and research talk participants who provided valuable feedback on this work. We also thank all collaborators and colleagues whose discussions and support helped improve the paper throughout its development.

\section*{Funding}
Y.-Y.A.\ acknowledges support from the U.S. National Science Foundation under Grant No.~2404109.
\section*{Data and Code Availability}

Data and Code will be available upon publication at \url{https://github.com/yulin-yu/datarepurpose} 

\section*{Supplementary Information}

This document includes:

Supplementary Note 1: Data Description

Supplementary Note 2: Example of Low and High data repurpose

Supplementary Note 3: Variable Description

Supplementary Note 4: Regression Tables

Supplementary Note 5: Robustness Check: Alternative Measures of Data Repurposing

\subsection*{1. Data Description}

We integrated three data resources to quantify the effect of data repurposing on scientific impact.

(1) SciSciNet dataset(version 1): a large-scale, open data lake for the science of science, integrating over 134 million scientific publications with rich external linkages such as funding records, patents, clinical trials, and media mentions. This dataset is built upon the Microsoft Academic Graph and extended with curated linkages and pre-computed scientometric indicators such as team size or novelty score.

(2) OpenAlex dataset: a fully open scientific knowledge graph, encompassing metadata for over 250 million works and 100 million authors. In this project, we used the OpenAlex dataset (OpenAlex API snapshot retrieved in August 2024) to extract publication information, including references and citations, since three-year citation counts for 2021 papers fall outside the range covered by the SciSciNet dataset.

(3) Papers with Code: an open repository linking scientific papers with their associated code implementations, benchmarks, and datasets across machine learning and related fields (June 16, 2021 snapshot). The dataset integrates metadata from sources such as arXiv and GitHub, enabling researchers to connect publications to runnable code and track benchmark performance over time. In this project, we used the Papers with Code dataset to identify datasets used in machine learning papers and link them to corresponding publications in SciSciNet and OpenAlex.

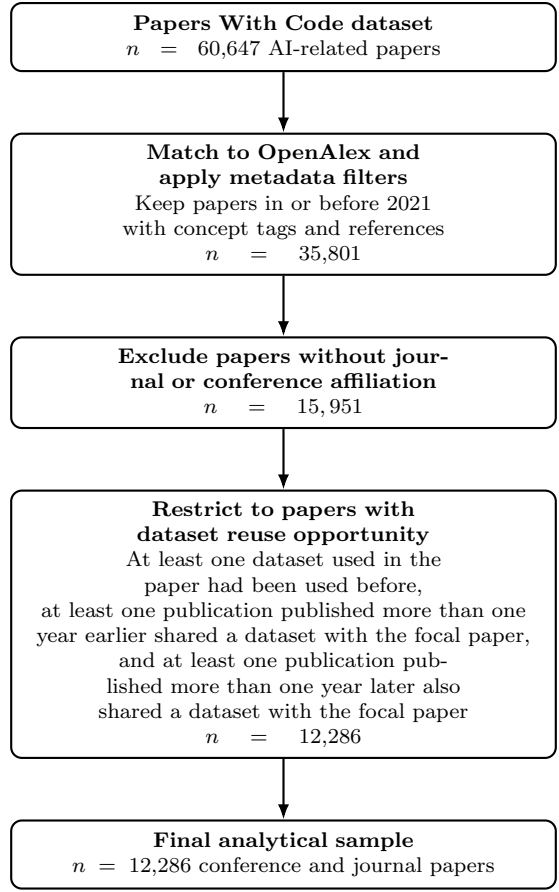
\begin{figure}[t]
\centering
\begin{tikzpicture}[
    node distance=0.8cm,
    every node/.style={font=\scriptsize, align=center},
    box/.style={
        rectangle,
        rounded corners,
        draw=black,
        thick,
        minimum width=7.2cm,
        minimum height=0.9cm,
        text width=6.8cm,
        inner sep=4pt
    },
    arrow/.style={
        -{Latex[length=2mm]},
        thick
    }
]

\node[box] (pwc) {\textbf{Papers With Code dataset}\\ $n=60{,}647$ AI-related papers};

\node[box, below=of pwc] (oa) {\textbf{Match to OpenAlex and apply metadata filters}\\ Keep papers in or before 2021 with concept tags and references\\ $n=35{,}801$};

\node[box, below=of oa] (m3) {\textbf{Exclude papers without journal or conference affiliation}\\  $n=15,951$};

\node[box, below=of m3] (m4) {\textbf{Restrict to papers with dataset reuse opportunity}\\ At least one dataset used in the paper had been used before,\\  at least one publication published more than one year earlier shared a dataset with the focal paper, \\ and at least one publication published more than one year later also shared a dataset with the focal paper
\\  $n=12{,}286$};

\node[box, below=of m4] (final) {\textbf{Final analytical sample}\\ $n=12{,}286$ conference and journal papers};

\draw[arrow] (pwc) -- (oa);
\draw[arrow] (oa) -- (m3);
\draw[arrow] (m3) -- (m4);
\draw[arrow] (m4) -- (final);

\end{tikzpicture}
\caption{Flow chart of sample construction.}
\label{fig:sample_flow}
\end{figure}

\subsection*{2. Example of Measurements}



| Prior Papers Using Dataset | Low Data Repurpose Score Papers | High Data Repurpose Score Papers |
|---|---|---|
| a remote sensing image dataset for cloud removal | cloud removal for remote sensing imagery via spatial attention generative adversarial network | rice grain disease identification using dual phase convolutional neural network based system aimed at small dataset |
| can neural networks understand monotonicity reasoning | combining event semantics and degree semantics for natural language inference | logical inferences with comparatives and generalized quantifiers |
| (1) 4dfab a large scale 4d database for facial expression analysis and biometric applications (2) meshgan non linear 3d morphable models of faces | detection of genuine and posed facial expressions of emotion a review | visual methods for sign language recognition a modality based review |
| a reduced dimension fmri shared response model | enabling factor analysis on thousand subject neuroimaging datasets | multi view treelet transform |
| sen12ms a curated dataset of georeferenced multi spectral sentinel 1 2 imagery for deep learning and data fusion | weakly supervised semantic segmentation of satellite images for land cover mapping challenges and opportunities | deep learning meets sar |
| audio tagging with noisy labels and minimal supervision | audio tagging by cross filtering noisy labels | fsd50k an open dataset of human labeled sound events |
| cosmoflow using deep learning to learn the universe at scale | the case for strong scaling in deep learning training large 3d cnns with hybrid parallelism | matrix engines for high performance computing a paragon of performance or grasping at straws |
| building sequential inference models for end to end response selection | dstc7 task 1 noetic end to end response selection | poly encoders transformer architectures and pre training strategies for fast and accurate multi sentence scoring |
| kannada mnist a new handwritten digits dataset for the kannada language | mnist mix a multi language handwritten digit recognition dataset | a random matrix analysis of random fourier features beyond the gaussian kernel a precise phase transition and the corresponding double descent |
| kannada mnist a new handwritten digits dataset for the kannada language | mnist mix a multi language handwritten digit recognition dataset | a random matrix analysis of random fourier features beyond the gaussian kernel a precise phase transition and the corresponding double descent |



\subsection*{3. Variable Descriptions}

{\bf Number of datasets:} The total number of datasets used in a paper.

\noindent {\bf 3 year Citation Counts:} Since SciSciNet only has citation count till 2022, we utilize the OpenAlex snapshot to extract all papers that have cited the targeted paper within a 3 year timeframe, starting from its publication year.

\noindent {\bf Publication year:} The year of publication is significantly associated with citation rates. Therefore, in order to control for potential time effects, we incorporate the year variable as a categorical variable. 

\noindent {\bf Dataset use Atypicality:}
Following \cite{yu2024does}, we quantify the atypicality of the datasets used by each focal paper $i$ with a Rao--Stirling--type index over its dataset set $D_i$. For every dataset $d$, we construct a binary usage vector $\mathbf{v}_d$ across the corpus and define the pairwise dissimilarity as cosine distance $\delta_{dd'} = 1 - \cos(\mathbf{v}_d,\mathbf{v}_{d'})$. Let $p_d$ denote the share of dataset $d$ within paper $i$ (for $K=|D_i|$, $p_d = 1/K$). The paper-level atypicality is
\[
\mathrm{Atyp}(i) \;=\; \sum_{d \neq d' \in D_i} p_d\, p_{d'}\, \delta_{dd'}.
\]
Higher values indicate more unusual (less frequently co-used) dataset combinations.

\noindent {\bf Dataset use frequency:} A paper utilizing a frequently used dataset may focus on popular research questions, which could potentially confound the citation analysis. To address this concern, we introduce dataset use frequency as a controlling variable when investigating its impact on citation rates. In cases where a paper incorporates multiple datasets, we calculate the mean number of papers that utilize datasets used by a focal paper.\\

\noindent {\bf Number of authors:} Research on team science suggests that the number of co-authors positively correlates with citation impact~\cite{uzzi2013atypical,wu2019large}, as a larger number of co-authors tends to result in a more extensive citation network.\\

\noindent {\bf Author Recognition:} Author recognition can serve as a proxy for experience and authority in the field. Moreover, this variable has a strong correlation with citation impact. We
measure author recognition of a paper as the average number of citations received by the
authors. In the regression analysis, we log-transform the Author Recognition variable using
Log(Author Recognition+0.01).\\


\noindent {\bf Journal Impact Factor:} The majority of journals in our dataset do not have a publicly recorded impact factor. To address this limitation, we employ an alternative approach by calculating the average citation count for papers published in these journals in 2019. This average citation count is used as a proxy for the impact factor of the journal. In the regression analysis, we log-transform the Journal Impact Factor variable using Log(Journal Impact Factor+0.01).

\noindent {\bf Number of Reference:} The total number of reference in a paper. 

\noindent {\bf Average Institution Impact:} Institution recognition can serve as a proxy for organizational prestige, resources, and network advantages, and it is often correlated with citation impact. We measure Institution Recognition for a paper as the average number of citations received by the institutions affiliated with its authors (averaged across the paper’s unique affiliations). In the regression analysis, we log-transform this variable as Log(Institution Recognition + 0.01).

\noindent {\bf Number of Institutions:} The total number of institutions listed in a paper, extracted from the SciSciNet dataset.
\subsection*{4. Regression Tables}
The full regression table is presented in Table 1 to Table 15.

\begin{figure}
\centering
\includegraphics[width=.6\linewidth]{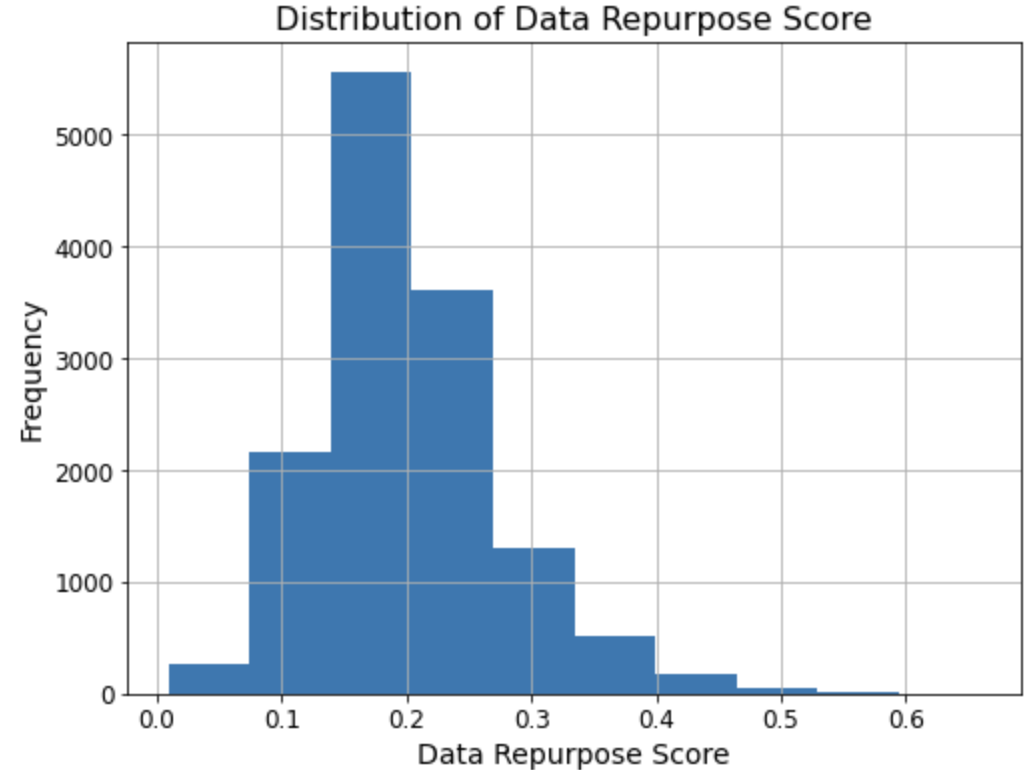}
\caption{Histogram of data repurposing scores.}
\label{fig:a5}
\end{figure}

\textbf{(1) The effect of dataset repurpose on D-Score (S1)} 
\FloatBarrier
\begin{table*}[ht]
\tiny
\begin{center}
  \renewcommand{\tablename}{Table S}

\begin{tabular}{lcccccc}
\textbf{Dep. Variable:} DisruptionScore & \textbf{coef} & \textbf{std err} & \textbf{t} & \textbf{P$>\,|t|$} & \textbf{[0.025} & \textbf{0.975]} \\
\hline
\hline
\textbf{Intercept}                           &  0.8232  & 0.112 &  7.366 & 0.000 &  0.604 &  1.042 \\
\textbf{C(Year)[T.2016.0]}                & -0.2164  & 0.059 & -3.681 & 0.000 & -0.332 & -0.101 \\
\textbf{C(Year)[T.2017.0]}                & -0.4691  & 0.055 & -8.607 & 0.000 & -0.576 & -0.362 \\
\textbf{C(Year)[T.2018.0]}                & -0.5360  & 0.051 & -10.564 & 0.000 & -0.635 & -0.437 \\
\textbf{C(Year)[T.2019.0]}                & -0.7886  & 0.050 & -15.781 & 0.000 & -0.887 & -0.691 \\
\textbf{C(Year)[T.2020.0]}                & -0.7044  & 0.049 & -14.300 & 0.000 & -0.801 & -0.608 \\
\textbf{C(Year)[T.2021.0]}                & -0.6920  & 0.060 & -11.588 & 0.000 & -0.809 & -0.575 \\
\textbf{C(Year)[T.before\_2015]}          &  0.5267  & 0.070 &  7.570 & 0.000 &  0.390 &  0.663 \\
\textbf{DataRecombination\_std}                &  0.0432  & 0.013 &  3.415 & 0.001 &  0.018 &  0.068 \\
\textbf{DataRepurposed\_std}            &  0.1022  & 0.011 &  9.648 & 0.000 &  0.081 &  0.123 \\
\textbf{PaperNovelty\_std}            & -0.2436  & 0.013 & -18.149 & 0.000 & -0.270 & -0.217 \\
\textbf{NumDataset\_log}                      & -0.1379  & 0.023 &  -6.005 & 0.000 & -0.183 & -0.093 \\
\textbf{data use frequency\_log}             &  0.0074  & 0.006 &   1.195 & 0.232 & -0.005 &  0.020 \\
\textbf{AuthorExperience\_log}                     & -0.0033  & 0.013 &  -0.245 & 0.806 & -0.030 &  0.023 \\
\textbf{ReferenceLen\_log}                 & -0.0089  & 0.023 &  -0.396 & 0.692 & -0.053 &  0.035 \\
\textbf{NumAuthor\_log}                     & -0.0341  & 0.026 &  -1.289 & 0.197 & -0.086 &  0.018 \\
\textbf{Citation\_log}                     &  0.1052  & 0.008 &  13.767 & 0.000 &  0.090 &  0.120 \\
\textbf{InstitutionImpact\_log} & -0.1384  & 0.028 &  -5.001 & 0.000 & -0.193 & -0.084 \\
\textbf{ImpactFactor\_log}                    & -0.0265  & 0.011 &  -2.354 & 0.019 & -0.049 & -0.004 \\
\textbf{NumInstitution}                       &  0.0081  & 0.011 &   0.758 & 0.448 & -0.013 &  0.029 \\
\hline
\textbf{No. Observations:} & 12286 & \textbf{R-squared:} & 0.143  & \textbf{Adj. R-squared:} & 0.142 & \\
\textbf{Df Residuals:}    & 12266 & \textbf{Df Model:}  & 19     & \textbf{Covariance Type:} & nonrobust & \\
\textbf{F-statistic:}     & 108.0 & \textbf{Prob(F-statistic):} & 0.00 & \textbf{AIC:} & 3.598e+04 & \\
\textbf{BIC:}             & 3.613e+04 & \textbf{Log-Likelihood:} & -17971. & \textbf{Durbin--Watson:} & 1.965 & \\
\textbf{Omnibus:}         & 1358.659 & \textbf{Prob(Omnibus):} & 0.000 & \textbf{Jarque--Bera (JB):} & 1939.350 & \\
\textbf{Prob(JB):}        & 0.00  & \textbf{Skew:} & -0.855 & \textbf{Kurtosis:} & 3.930 & \\
\textbf{Cond. No.:}       & 133.  &  &  &  &  & \\
\end{tabular}

\caption{OLS regression results with \texttt{DisruptionScore} as the dependent variable and data repurposing as the key predictor.}
\end{center}
\end{table*}

\textbf{(2) The effect of dataset repurpose on Citation (S2)}
\FloatBarrier
\begin{table*}[ht]
\tiny
\begin{center}
  \renewcommand{\tablename}{Table S}

\begin{tabular}{lcccccc}
\textbf{Dep. Variable:} Citation & \textbf{coef} & \textbf{std err} & \textbf{z} & \textbf{P$>\,|z|$} & \textbf{[0.025} & \textbf{0.975]} \\
\hline
\hline
\textbf{Intercept}                           & 1.3287 & 0.082 & 16.283 & 0.000 & 1.169 & 1.489 \\
\textbf{C(Year)[T.2016.0]}                & 0.0842 & 0.057 &  1.487 & 0.137 & -0.027 & 0.195 \\
\textbf{C(Year)[T.2017.0]}                & 0.2259 & 0.052 &  4.308 & 0.000 & 0.123 & 0.329 \\
\textbf{C(Year)[T.2018.0]}                & 0.0378 & 0.049 &  0.774 & 0.439 & -0.058 & 0.134 \\
\textbf{C(Year)[T.2019.0]}                & -0.1783 & 0.048 & -3.709 & 0.000 & -0.273 & -0.084 \\
\textbf{C(Year)[T.2020.0]}                & -0.3818 & 0.047 & -8.078 & 0.000 & -0.474 & -0.289 \\
\textbf{C(Year)[T.2021.0]}                & -0.7684 & 0.057 & -13.490 & 0.000 & -0.880 & -0.657 \\
\textbf{C(Year)[T.before\_2015]}          & -0.1047 & 0.067 & -1.561 & 0.119 & -0.236 & 0.027 \\
\textbf{DataRecombination\_std}                & 0.1400 & 0.012 & 11.505 & 0.000 & 0.116 & 0.164 \\
\textbf{DataRepurposed\_std}            & -0.0084 & 0.010 & -0.820 & 0.412 & -0.028 & 0.012 \\
\textbf{PaperNovelty\_std}            & 0.0766 & 0.013 &  6.024 & 0.000 & 0.052 & 0.102 \\
\textbf{NumDataset\_log}                      & 0.5861 & 0.022 & 26.939 & 0.000 & 0.543 & 0.629 \\
\textbf{data use frequency\_log}             & 0.0263 & 0.006 &  4.390 & 0.000 & 0.015 & 0.038 \\
\textbf{AuthorExperience\_log}                     & 0.1101 & 0.013 &  8.511 & 0.000 & 0.085 & 0.135 \\
\textbf{NumAuthor\_log}                     & 0.3557 & 0.025 & 14.082 & 0.000 & 0.306 & 0.405 \\
\textbf{InstitutionImpact\_log} & 0.5117 & 0.026 & 19.335 & 0.000 & 0.460 & 0.564 \\
\textbf{ImpactFactor\_log}                    & 0.3661 & 0.010 & 35.298 & 0.000 & 0.346 & 0.386 \\
\textbf{NumInstitution}                       & 0.0152 & 0.010 &  1.493 & 0.135 & -0.005 & 0.035 \\
\hline
\textbf{No. Observations:} & 12286 & \textbf{Model:} & GLM & \textbf{Method:} & IRLS & \\
\textbf{Model Family:} & NegativeBinomial & \textbf{Link Function:} & Log & \textbf{Covariance Type:} & nonrobust & \\
\textbf{Df Residuals:} & 12268 & \textbf{Df Model:} & 17 & \textbf{No. Iterations:} & 17 & \\
\textbf{Log-Likelihood:} & -65186. & \textbf{Deviance:} & 19039. & \textbf{Pearson chi2:} & 5.06e+04 & \\
\textbf{Pseudo R-squ. (CS):} & 0.4251 & & & & & \\
\end{tabular}

\caption{Negative binomial GLM with \texttt{Citation} as the dependent variable.}
\end{center}
\end{table*}

\textbf{(3) The effect of repurpose propagation on D-Score (S3)}
\FloatBarrier
\begin{table*}[ht]
\tiny
\begin{center}
  \renewcommand{\tablename}{Table S}

\begin{tabular}{lcccccc}
\textbf{Dep. Variable:} DisruptionScore& \textbf{coef} & \textbf{std err} & \textbf{t} & \textbf{P$>\,|t|$} & \textbf{[0.025} & \textbf{0.975]} \\
\hline
\hline
\textbf{Intercept}                           & 0.6571 & 0.111 &  5.931 & 0.000 & 0.440 & 0.874 \\
\textbf{C(Year)[T.2016.0]}                & -0.2172 & 0.059 & -3.681 & 0.000 & -0.333 & -0.102 \\
\textbf{C(Year)[T.2017.0]}                & -0.4693 & 0.055 & -8.579 & 0.000 & -0.576 & -0.362 \\
\textbf{C(Year)[T.2018.0]}                & -0.5328 & 0.051 & -10.464 & 0.000 & -0.633 & -0.433 \\
\textbf{C(Year)[T.2019.0]}                & -0.7793 & 0.050 & -15.541 & 0.000 & -0.878 & -0.681 \\
\textbf{C(Year)[T.2020.0]}                & -0.6959 & 0.049 & -14.078 & 0.000 & -0.793 & -0.599 \\
\textbf{C(Year)[T.2021.0]}                & -0.6875 & 0.061 & -11.324 & 0.000 & -0.806 & -0.568 \\
\textbf{C(Year)[T.before\_2015]}          & 0.5358 & 0.070 &  7.671 & 0.000 & 0.399 & 0.673 \\
\textbf{DataRecombination\_std}                & 0.0530 & 0.013 &  4.182 & 0.000 & 0.028 & 0.078 \\
\textbf{RepurposePropagation\_std}          & 0.0285 & 0.010 &  2.939 & 0.003 & 0.010 & 0.048 \\
\textbf{PaperNovelty\_std}            & -0.2692 & 0.013 & -20.403 & 0.000 & -0.295 & -0.243 \\
\textbf{NumDataset\_log}                      & -0.1262 & 0.023 & -5.474 & 0.000 & -0.171 & -0.081 \\
\textbf{data use frequency\_log}             & 0.0256 & 0.006 &  4.264 & 0.000 & 0.014 & 0.037 \\
\textbf{AuthorExperience\_log}                     & 0.0049 & 0.013 &  0.361 & 0.718 & -0.022 & 0.031 \\
\textbf{ReferenceLen\_log}                 & -0.0039 & 0.023 & -0.172 & 0.864 & -0.048 & 0.040 \\
\textbf{NumAuthor\_log}                     & -0.0293 & 0.027 & -1.105 & 0.269 & -0.081 & 0.023 \\
\textbf{Citation\_log}                     & 0.0980 & 0.008 & 12.790 & 0.000 & 0.083 & 0.113 \\
\textbf{InstitutionImpact\_log} & -0.1203 & 0.028 & -4.344 & 0.000 & -0.175 & -0.066 \\
\textbf{ImpactFactor\_log}                    & -0.0306 & 0.011 & -2.708 & 0.007 & -0.053 & -0.008 \\
\textbf{NumInstitution}                       & 0.0089 & 0.011 &  0.832 & 0.405 & -0.012 & 0.030 \\
\hline
\textbf{No. Observations:} & 12286 & \textbf{R-squared:} & 0.137 & \textbf{Adj. R-squared:} & 0.136 & \\
\textbf{Df Residuals:}    & 12266 & \textbf{Df Model:}  & 19    & \textbf{Covariance Type:} & nonrobust & \\
\textbf{F-statistic:}     & 102.9 & \textbf{Prob(F-statistic):} & 0.00 & \textbf{AIC:} & 3.607e+04 & \\
\textbf{BIC:}             & 3.621e+04 & \textbf{Log-Likelihood:} & -18013. & \textbf{Durbin--Watson:} & 1.967 & \\
\textbf{Omnibus:}         & 1405.594 & \textbf{Prob(Omnibus):} & 0.000 & \textbf{Jarque--Bera (JB):} & 2028.460 & \\
\textbf{Prob(JB):}        & 0.00  & \textbf{Skew:} & -0.873 & \textbf{Kurtosis:} & 3.956 & \\
\textbf{Cond. No.:}       & 133.  &  &  &  &  & \\
\end{tabular}

\caption{OLS regression results with \texttt{DisruptionScore} as the dependent variable and repurpose propagation as the key predictor.}
\end{center}
\end{table*}

\textbf{(4) The effect of repurpose propagation on Citation (S4)}
\FloatBarrier
\begin{table*}[ht]
\tiny
\begin{center}
  \renewcommand{\tablename}{Table S}

\begin{tabular}{lcccccc}
\textbf{Dep. Variable:} Citation & \textbf{coef} & \textbf{std err} & \textbf{z} & \textbf{P$>\,|z|$} & \textbf{[0.025} & \textbf{0.975]} \\
\hline
\hline
\textbf{Intercept}                           & 1.3733 & 0.080 & 17.143 & 0.000 & 1.216 & 1.530 \\
\textbf{C(Year)[T.2016.0]}                & 0.0683 & 0.057 &  1.207 & 0.228 & -0.043 & 0.179 \\
\textbf{C(Year)[T.2017.0]}                & 0.1999 & 0.052 &  3.813 & 0.000 & 0.097 & 0.303 \\
\textbf{C(Year)[T.2018.0]}                & 0.0108 & 0.049 &  0.222 & 0.825 & -0.085 & 0.107 \\
\textbf{C(Year)[T.2019.0]}                & -0.2138 & 0.048 & -4.447 & 0.000 & -0.308 & -0.120 \\
\textbf{C(Year)[T.2020.0]}                & -0.4136 & 0.047 & -8.753 & 0.000 & -0.506 & -0.321 \\
\textbf{C(Year)[T.2021.0]}                & -0.8698 & 0.058 & -15.094 & 0.000 & -0.983 & -0.757 \\
\textbf{C(Year)[T.before\_2015]}          & -0.1322 & 0.067 & -1.971 & 0.049 & -0.264 & -0.001 \\
\textbf{DataRecombination\_std}                & 0.1400 & 0.012 & 11.534 & 0.000 & 0.116 & 0.164 \\
\textbf{RepurposePropagation\_std}          & 0.0831 & 0.009 &  8.910 & 0.000 & 0.065 & 0.101 \\
\textbf{PaperNovelty\_std}            & 0.0813 & 0.012 &  6.530 & 0.000 & 0.057 & 0.106 \\
\textbf{NumDataset\_log}                      & 0.5870 & 0.022 & 26.972 & 0.000 & 0.544 & 0.630 \\
\textbf{data use frequency\_log}             & 0.0265 & 0.006 &  4.588 & 0.000 & 0.015 & 0.038 \\
\textbf{AuthorExperience\_log}                     & 0.1082 & 0.013 &  8.377 & 0.000 & 0.083 & 0.133 \\
\textbf{NumAuthor\_log}                     & 0.3645 & 0.025 & 14.428 & 0.000 & 0.315 & 0.414 \\
\textbf{InstitutionImpact\_log} & 0.5001 & 0.026 & 18.927 & 0.000 & 0.448 & 0.552 \\
\textbf{ImpactFactor\_log}                    & 0.3669 & 0.010 & 35.426 & 0.000 & 0.347 & 0.387 \\
\textbf{NumInstitution}                       & 0.0161 & 0.010 &  1.581 & 0.114 & -0.004 & 0.036 \\
\hline
\textbf{No. Observations:} & 12286 & \textbf{Model:} & GLM & \textbf{Method:} & IRLS & \\
\textbf{Model Family:} & NegativeBinomial & \textbf{Link Function:} & Log & \textbf{Covariance Type:} & nonrobust & \\
\textbf{Df Residuals:} & 12268 & \textbf{Df Model:} & 17 & \textbf{No. Iterations:} & 16 & \\
\textbf{Log-Likelihood:} & -65144. & \textbf{Deviance:} & 18954. & \textbf{Pearson chi2:} & 5.03e+04 & \\
\textbf{Pseudo R-squ. (CS):} & 0.4290 & & & & & \\
\end{tabular}

\caption{Negative binomial GLM with \texttt{Citation} as the dependent variable.}
\end{center}
\end{table*}

\newpage

\textbf{Team-characteristic models for Figure~\ref{fig:a4} (S5--S14)}

Tables~S5--S14 report the OLS models underlying Figure~\ref{fig:a4}, estimated on the same 12,286-paper full regression sample used in Tables~S1--S4. Each model includes one team-characteristic predictor and controls for dataset-use atypicality, number of datasets, and dataset-use frequency.

\textbf{(5) Team Characters and data repurposing (S5-S9)}

\FloatBarrier
\begin{table*}[ht]
\tiny
\begin{center}
  \renewcommand{\tablename}{Table S}

\begin{tabular}{lcccccc}
\textbf{Dep. Variable:} DataRepurposing\_std & \textbf{coef} & \textbf{std err} & \textbf{t} & \textbf{P$>\,|t|$} & \textbf{[0.025} & \textbf{0.975]} \\
\hline
\hline
\textbf{Intercept} & -1.0352 & 0.040 & -26.170 & 0.000 & -1.1128 & -0.9577 \\
\textbf{recomb\_standardized} & 0.1046 & 0.011 & 9.397 & 0.000 & 0.0828 & 0.1264 \\
\textbf{num\_data\_log} & 0.0143 & 0.019 & 0.735 & 0.462 & -0.0237 & 0.0523 \\
\textbf{datause\_frequency\_log} & 0.1579 & 0.005 & 31.572 & 0.000 & 0.1481 & 0.1677 \\
\textbf{Team\_Size\_log} & 0.0449 & 0.020 & 2.275 & 0.023 & 0.0062 & 0.0835 \\
\hline
\textbf{No. Observations:} & 12286 & \textbf{R-squared:} & 0.082 & \textbf{Adj. R-squared:} & 0.081 & \\
\textbf{Df Residuals:} & 12281 & \textbf{Df Model:} & 4 & \textbf{Covariance Type:} & nonrobust & \\
\textbf{F-statistic:} & 273.489 & \textbf{Prob(F-statistic):} & 0.000 & \textbf{AIC:} & 3.3e+04 & \\
\textbf{BIC:} & 3.3e+04 & \textbf{Log-Likelihood:} & -1.65e+04 &  &  & \\
\end{tabular}

\caption{OLS regression results for Team size and data repurposing. The model controls for \texttt{DataRecombination\_std}, \texttt{NumDataset\_log}, and \texttt{data use frequency\_log}. The model is estimated on the same 12,286-paper full regression sample used in Tables~S1--S4.}
\end{center}
\end{table*}

\begin{table*}[ht]
\tiny
\begin{center}
  \renewcommand{\tablename}{Table S}

\begin{tabular}{lcccccc}
\textbf{Dep. Variable:} DataRepurposing\_std & \textbf{coef} & \textbf{std err} & \textbf{t} & \textbf{P$>\,|t|$} & \textbf{[0.025} & \textbf{0.975]} \\
\hline
\hline
\textbf{Intercept} & -1.0763 & 0.040 & -26.894 & 0.000 & -1.1548 & -0.9979 \\
\textbf{recomb\_standardized} & 0.1041 & 0.011 & 9.354 & 0.000 & 0.0823 & 0.1259 \\
\textbf{num\_data\_log} & 0.0156 & 0.019 & 0.805 & 0.421 & -0.0223 & 0.0534 \\
\textbf{datause\_frequency\_log} & 0.1576 & 0.005 & 31.509 & 0.000 & 0.1478 & 0.1674 \\
\textbf{Average\_LogC10} & 0.0410 & 0.011 & 3.755 & 0.000 & 0.0196 & 0.0623 \\
\hline
\textbf{No. Observations:} & 12286 & \textbf{R-squared:} & 0.082 & \textbf{Adj. R-squared:} & 0.082 & \\
\textbf{Df Residuals:} & 12281 & \textbf{Df Model:} & 4 & \textbf{Covariance Type:} & nonrobust & \\
\textbf{F-statistic:} & 275.918 & \textbf{Prob(F-statistic):} & 0.000 & \textbf{AIC:} & 3.3e+04 & \\
\textbf{BIC:} & 3.3e+04 & \textbf{Log-Likelihood:} & -1.65e+04 &  &  & \\
\end{tabular}

\caption{OLS regression results for Team experience and data repurposing. The model controls for \texttt{DataRecombination\_std}, \texttt{NumDataset\_log}, and \texttt{data use frequency\_log}. The model is estimated on the same 12,286-paper full regression sample used in Tables~S1--S4.}
\end{center}
\end{table*}

\begin{table*}[ht]
\tiny
\begin{center}
  \renewcommand{\tablename}{Table S}

\begin{tabular}{lcccccc}
\textbf{Dep. Variable:} DataRepurposing\_std & \textbf{coef} & \textbf{std err} & \textbf{t} & \textbf{P$>\,|t|$} & \textbf{[0.025} & \textbf{0.975]} \\
\hline
\hline
\textbf{Intercept} & -1.1783 & 0.053 & -22.093 & 0.000 & -1.2828 & -1.0737 \\
\textbf{recomb\_standardized} & 0.1036 & 0.011 & 9.305 & 0.000 & 0.0817 & 0.1254 \\
\textbf{num\_data\_log} & 0.0119 & 0.019 & 0.616 & 0.538 & -0.0260 & 0.0499 \\
\textbf{datause\_frequency\_log} & 0.1582 & 0.005 & 31.651 & 0.000 & 0.1484 & 0.1680 \\
\textbf{avg\_inst\_impact\_since\_2019\_log} & 0.1014 & 0.022 & 4.571 & 0.000 & 0.0579 & 0.1449 \\
\hline
\textbf{No. Observations:} & 12286 & \textbf{R-squared:} & 0.083 & \textbf{Adj. R-squared:} & 0.083 & \\
\textbf{Df Residuals:} & 12281 & \textbf{Df Model:} & 4 & \textbf{Covariance Type:} & nonrobust & \\
\textbf{F-statistic:} & 277.767 & \textbf{Prob(F-statistic):} & 0.000 & \textbf{AIC:} & 3.3e+04 & \\
\textbf{BIC:} & 3.3e+04 & \textbf{Log-Likelihood:} & -1.65e+04 &  &  & \\
\end{tabular}

\caption{OLS regression results for Institutional impact and data repurposing. The model controls for \texttt{DataRecombination\_std}, \texttt{NumDataset\_log}, and \texttt{data use frequency\_log}. The model is estimated on the same 12,286-paper full regression sample used in Tables~S1--S4.}
\end{center}
\end{table*}

\begin{table*}[ht]
\tiny
\begin{center}
  \renewcommand{\tablename}{Table S}

\begin{tabular}{lcccccc}
\textbf{Dep. Variable:} DataRepurposing\_std & \textbf{coef} & \textbf{std err} & \textbf{t} & \textbf{P$>\,|t|$} & \textbf{[0.025} & \textbf{0.975]} \\
\hline
\hline
\textbf{Intercept} & -1.1992 & 0.036 & -33.691 & 0.000 & -1.2689 & -1.1294 \\
\textbf{C(is\_developed\_country)[T.mixed]} & 0.1909 & 0.026 & 7.470 & 0.000 & 0.1408 & 0.2410 \\
\textbf{C(is\_developed\_country)[T.true]} & 0.3146 & 0.024 & 12.892 & 0.000 & 0.2667 & 0.3624 \\
\textbf{recomb\_standardized} & 0.0995 & 0.011 & 8.990 & 0.000 & 0.0778 & 0.1212 \\
\textbf{num\_data\_log} & 0.0169 & 0.019 & 0.878 & 0.380 & -0.0208 & 0.0545 \\
\textbf{datause\_frequency\_log} & 0.1580 & 0.005 & 31.799 & 0.000 & 0.1482 & 0.1677 \\
\hline
\textbf{No. Observations:} & 12286 & \textbf{R-squared:} & 0.094 & \textbf{Adj. R-squared:} & 0.094 & \\
\textbf{Df Residuals:} & 12280 & \textbf{Df Model:} & 5 & \textbf{Covariance Type:} & nonrobust & \\
\textbf{F-statistic:} & 255.647 & \textbf{Prob(F-statistic):} & 0.000 & \textbf{AIC:} & 3.28e+04 & \\
\textbf{BIC:} & 3.29e+04 & \textbf{Log-Likelihood:} & -1.64e+04 &  &  & \\
\end{tabular}

\caption{OLS regression results for Country development status and data repurposing. The model controls for \texttt{DataRecombination\_std}, \texttt{NumDataset\_log}, and \texttt{data use frequency\_log}. The model is estimated on the same 12,286-paper full regression sample used in Tables~S1--S4.}
\end{center}
\end{table*}

\begin{table*}[ht]
\tiny
\begin{center}
  \renewcommand{\tablename}{Table S}

\begin{tabular}{lcccccc}
\textbf{Dep. Variable:} DataRepurposing\_std & \textbf{coef} & \textbf{std err} & \textbf{t} & \textbf{P$>\,|t|$} & \textbf{[0.025} & \textbf{0.975]} \\
\hline
\hline
\textbf{Intercept} & -0.9939 & 0.031 & -32.331 & 0.000 & -1.0541 & -0.9336 \\
\textbf{C(academic\_or\_industry)[T.industry]} & 0.0042 & 0.034 & 0.124 & 0.901 & -0.0620 & 0.0704 \\
\textbf{C(academic\_or\_industry)[T.mixed]} & 0.0452 & 0.018 & 2.580 & 0.010 & 0.0109 & 0.0796 \\
\textbf{recomb\_standardized} & 0.1049 & 0.011 & 9.425 & 0.000 & 0.0831 & 0.1267 \\
\textbf{num\_data\_log} & 0.0161 & 0.019 & 0.833 & 0.405 & -0.0218 & 0.0540 \\
\textbf{datause\_frequency\_log} & 0.1581 & 0.005 & 31.608 & 0.000 & 0.1483 & 0.1679 \\
\hline
\textbf{No. Observations:} & 12286 & \textbf{R-squared:} & 0.082 & \textbf{Adj. R-squared:} & 0.082 & \\
\textbf{Df Residuals:} & 12280 & \textbf{Df Model:} & 5 & \textbf{Covariance Type:} & nonrobust & \\
\textbf{F-statistic:} & 219.141 & \textbf{Prob(F-statistic):} & 0.000 & \textbf{AIC:} & 3.3e+04 & \\
\textbf{BIC:} & 3.3e+04 & \textbf{Log-Likelihood:} & -1.65e+04 &  &  & \\
\end{tabular}

\caption{OLS regression results for Sectoral background and data repurposing. The model controls for \texttt{DataRecombination\_std}, \texttt{NumDataset\_log}, and \texttt{data use frequency\_log}. The model is estimated on the same 12,286-paper full regression sample used in Tables~S1--S4.}
\end{center}
\end{table*}

\textbf{(6) Team size and repurpose propagation (S10-15)}
\FloatBarrier
\begin{table*}[ht]
\tiny
\begin{center}
  \renewcommand{\tablename}{Table S}

\begin{tabular}{lcccccc}
\textbf{Dep. Variable:} RepurposePropagation\_std & \textbf{coef} & \textbf{std err} & \textbf{t} & \textbf{P$>\,|t|$} & \textbf{[0.025} & \textbf{0.975]} \\
\hline
\hline
\textbf{Intercept} & -0.0620 & 0.043 & -1.436 & 0.151 & -0.1466 & 0.0226 \\
\textbf{recomb\_standardized} & -0.0020 & 0.012 & -0.165 & 0.869 & -0.0258 & 0.0218 \\
\textbf{num\_data\_log} & 0.0088 & 0.021 & 0.418 & 0.676 & -0.0326 & 0.0503 \\
\textbf{datause\_frequency\_log} & -0.0119 & 0.005 & -2.183 & 0.029 & -0.0226 & -0.0012 \\
\textbf{Team\_Size\_log} & 0.0661 & 0.022 & 3.072 & 0.002 & 0.0239 & 0.1082 \\
\hline
\textbf{No. Observations:} & 12286 & \textbf{R-squared:} & 0.001 & \textbf{Adj. R-squared:} & 0.000859 & \\
\textbf{Df Residuals:} & 12281 & \textbf{Df Model:} & 4 & \textbf{Covariance Type:} & nonrobust & \\
\textbf{F-statistic:} & 3.640 & \textbf{Prob(F-statistic):} & 0.006 & \textbf{AIC:} & 3.51e+04 & \\
\textbf{BIC:} & 3.52e+04 & \textbf{Log-Likelihood:} & -1.76e+04 &  &  & \\
\end{tabular}

\caption{OLS regression results for Team size and repurpose propagation. The model controls for \texttt{DataRecombination\_std}, \texttt{NumDataset\_log}, and \texttt{data use frequency\_log}. The model is estimated on the same 12,286-paper full regression sample used in Tables~S1--S4.}
\end{center}
\end{table*}

\begin{table*}[ht]
\tiny
\begin{center}
  \renewcommand{\tablename}{Table S}

\begin{tabular}{lcccccc}
\textbf{Dep. Variable:} RepurposePropagation\_std & \textbf{coef} & \textbf{std err} & \textbf{t} & \textbf{P$>\,|t|$} & \textbf{[0.025} & \textbf{0.975]} \\
\hline
\hline
\textbf{Intercept} & 0.0299 & 0.044 & 0.685 & 0.493 & -0.0557 & 0.1156 \\
\textbf{recomb\_standardized} & -0.0015 & 0.012 & -0.120 & 0.905 & -0.0253 & 0.0224 \\
\textbf{num\_data\_log} & 0.0142 & 0.021 & 0.674 & 0.500 & -0.0271 & 0.0556 \\
\textbf{datause\_frequency\_log} & -0.0117 & 0.005 & -2.141 & 0.032 & -0.0224 & -0.000986 \\
\textbf{Average\_LogC10} & -0.0020 & 0.012 & -0.170 & 0.865 & -0.0254 & 0.0213 \\
\hline
\textbf{No. Observations:} & 12286 & \textbf{R-squared:} & 0.000419 & \textbf{Adj. R-squared:} & 9.36e-05 & \\
\textbf{Df Residuals:} & 12281 & \textbf{Df Model:} & 4 & \textbf{Covariance Type:} & nonrobust & \\
\textbf{F-statistic:} & 1.288 & \textbf{Prob(F-statistic):} & 0.272 & \textbf{AIC:} & 3.51e+04 & \\
\textbf{BIC:} & 3.52e+04 & \textbf{Log-Likelihood:} & -1.76e+04 &  &  & \\
\end{tabular}

\caption{OLS regression results for Team experience and repurpose propagation. The model controls for \texttt{DataRecombination\_std}, \texttt{NumDataset\_log}, and \texttt{data use frequency\_log}. The model is estimated on the same 12,286-paper full regression sample used in Tables~S1--S4.}
\end{center}
\end{table*}

\textbf{(6) Institutional impact and repurpose propagation (S12-15)}
\FloatBarrier
\begin{table*}[ht]
\tiny
\begin{center}
  \renewcommand{\tablename}{Table S}

\begin{tabular}{lcccccc}
\textbf{Dep. Variable:} RepurposePropagation\_std & \textbf{coef} & \textbf{std err} & \textbf{t} & \textbf{P$>\,|t|$} & \textbf{[0.025} & \textbf{0.975]} \\
\hline
\hline
\textbf{Intercept} & -0.0952 & 0.058 & -1.636 & 0.102 & -0.2094 & 0.0189 \\
\textbf{recomb\_standardized} & -0.0023 & 0.012 & -0.192 & 0.848 & -0.0261 & 0.0215 \\
\textbf{num\_data\_log} & 0.0106 & 0.021 & 0.501 & 0.616 & -0.0308 & 0.0520 \\
\textbf{datause\_frequency\_log} & -0.0116 & 0.005 & -2.130 & 0.033 & -0.0223 & -0.000925 \\
\textbf{avg\_inst\_impact\_since\_2019\_log} & 0.0603 & 0.024 & 2.491 & 0.013 & 0.0129 & 0.1078 \\
\hline
\textbf{No. Observations:} & 12286 & \textbf{R-squared:} & 0.000922 & \textbf{Adj. R-squared:} & 0.000596 & \\
\textbf{Df Residuals:} & 12281 & \textbf{Df Model:} & 4 & \textbf{Covariance Type:} & nonrobust & \\
\textbf{F-statistic:} & 2.833 & \textbf{Prob(F-statistic):} & 0.023 & \textbf{AIC:} & 3.51e+04 & \\
\textbf{BIC:} & 3.52e+04 & \textbf{Log-Likelihood:} & -1.76e+04 &  &  & \\
\end{tabular}

\caption{OLS regression results for Institutional impact and repurpose propagation. The model controls for \texttt{DataRecombination\_std}, \texttt{NumDataset\_log}, and \texttt{data use frequency\_log}. The model is estimated on the same 12,286-paper full regression sample used in Tables~S1--S4.}
\end{center}
\end{table*}

\begin{table*}[ht]
\tiny
\begin{center}
  \renewcommand{\tablename}{Table S}

\begin{tabular}{lcccccc}
\textbf{Dep. Variable:} RepurposePropagation\_std & \textbf{coef} & \textbf{std err} & \textbf{t} & \textbf{P$>\,|t|$} & \textbf{[0.025} & \textbf{0.975]} \\
\hline
\hline
\textbf{Intercept} & 0.0473 & 0.039 & 1.210 & 0.226 & -0.0293 & 0.1240 \\
\textbf{C(is\_developed\_country)[T.mixed]} & -0.0215 & 0.028 & -0.768 & 0.443 & -0.0766 & 0.0335 \\
\textbf{C(is\_developed\_country)[T.true]} & -0.0298 & 0.027 & -1.111 & 0.267 & -0.0823 & 0.0228 \\
\textbf{recomb\_standardized} & -0.0010 & 0.012 & -0.083 & 0.934 & -0.0248 & 0.0228 \\
\textbf{num\_data\_log} & 0.0143 & 0.021 & 0.677 & 0.498 & -0.0271 & 0.0556 \\
\textbf{datause\_frequency\_log} & -0.0117 & 0.005 & -2.146 & 0.032 & -0.0224 & -0.0010 \\
\hline
\textbf{No. Observations:} & 12286 & \textbf{R-squared:} & 0.000518 & \textbf{Adj. R-squared:} & 0.000111 & \\
\textbf{Df Residuals:} & 12280 & \textbf{Df Model:} & 5 & \textbf{Covariance Type:} & nonrobust & \\
\textbf{F-statistic:} & 1.272 & \textbf{Prob(F-statistic):} & 0.273 & \textbf{AIC:} & 3.51e+04 & \\
\textbf{BIC:} & 3.52e+04 & \textbf{Log-Likelihood:} & -1.76e+04 &  &  & \\
\end{tabular}

\caption{OLS regression results for Country development status and repurpose propagation. The model controls for \texttt{DataRecombination\_std}, \texttt{NumDataset\_log}, and \texttt{data use frequency\_log}. The model is estimated on the same 12,286-paper full regression sample used in Tables~S1--S4.}
\end{center}
\end{table*}

\begin{table*}[ht]
\tiny
\begin{center}
  \renewcommand{\tablename}{Table S}

\begin{tabular}{lcccccc}
\textbf{Dep. Variable:} RepurposePropagation\_std & \textbf{coef} & \textbf{std err} & \textbf{t} & \textbf{P$>\,|t|$} & \textbf{[0.025} & \textbf{0.975]} \\
\hline
\hline
\textbf{Intercept} & 0.0072 & 0.034 & 0.216 & 0.829 & -0.0585 & 0.0730 \\
\textbf{C(academic\_or\_industry)[T.industry]} & 0.0298 & 0.037 & 0.809 & 0.418 & -0.0424 & 0.1021 \\
\textbf{C(academic\_or\_industry)[T.mixed]} & 0.0421 & 0.019 & 2.201 & 0.028 & 0.0046 & 0.0796 \\
\textbf{recomb\_standardized} & -0.0017 & 0.012 & -0.136 & 0.892 & -0.0255 & 0.0222 \\
\textbf{num\_data\_log} & 0.0123 & 0.021 & 0.583 & 0.560 & -0.0291 & 0.0537 \\
\textbf{datause\_frequency\_log} & -0.0117 & 0.005 & -2.149 & 0.032 & -0.0224 & -0.0010 \\
\hline
\textbf{No. Observations:} & 12286 & \textbf{R-squared:} & 0.00082 & \textbf{Adj. R-squared:} & 0.000413 & \\
\textbf{Df Residuals:} & 12280 & \textbf{Df Model:} & 5 & \textbf{Covariance Type:} & nonrobust & \\
\textbf{F-statistic:} & 2.015 & \textbf{Prob(F-statistic):} & 0.073 & \textbf{AIC:} & 3.51e+04 & \\
\textbf{BIC:} & 3.52e+04 & \textbf{Log-Likelihood:} & -1.76e+04 &  &  & \\
\end{tabular}

\caption{OLS regression results for Sectoral background and repurpose propagation. The model controls for \texttt{DataRecombination\_std}, \texttt{NumDataset\_log}, and \texttt{data use frequency\_log}. The model is estimated on the same 12,286-paper full regression sample used in Tables~S1--S4.}
\end{center}
\end{table*}

\newpage
\subsection*{5. Robustness Check: Alternative Measures of Data Repurposing}

\begin{figure}[tbhp]
\centering
\includegraphics[width=.9\linewidth]{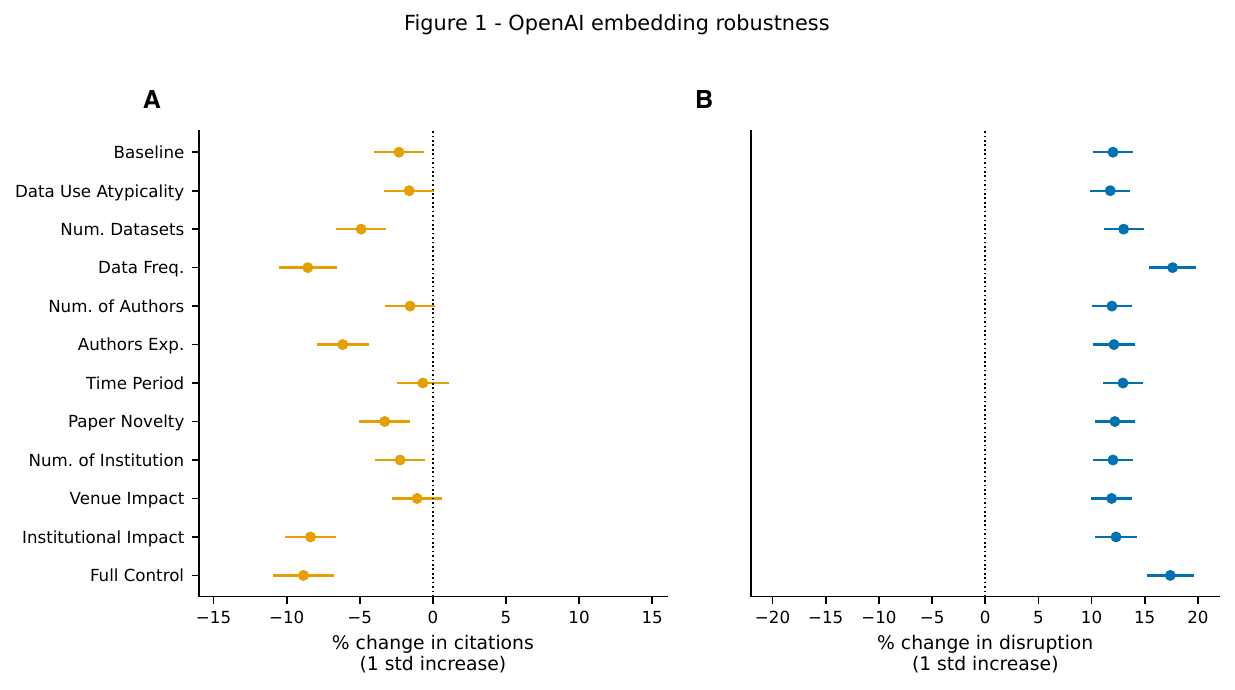}
\caption{
Data repurposing is associated with scientific disruption but not with short-term citation gains.
(A) Effect size of \textbf{data repurposing} on \textbf{citation count} over three years, estimated using negative binomial regressions. Each intermediate row reports the data repurposing coefficient from a separate regression that includes data repurposing and the control variable(s) indicated in that row. The baseline model includes no control variables, whereas the full model includes all control variables simultaneously. 
(B) Effect size of \textbf{data repurposing} on the disruption score over three years, estimated using OLS regressions with the same control structure as in (A).
}
\label{fig:a1_openai}
\end{figure}

\begin{figure}[tbhp]
\centering
\includegraphics[width=.9\linewidth]{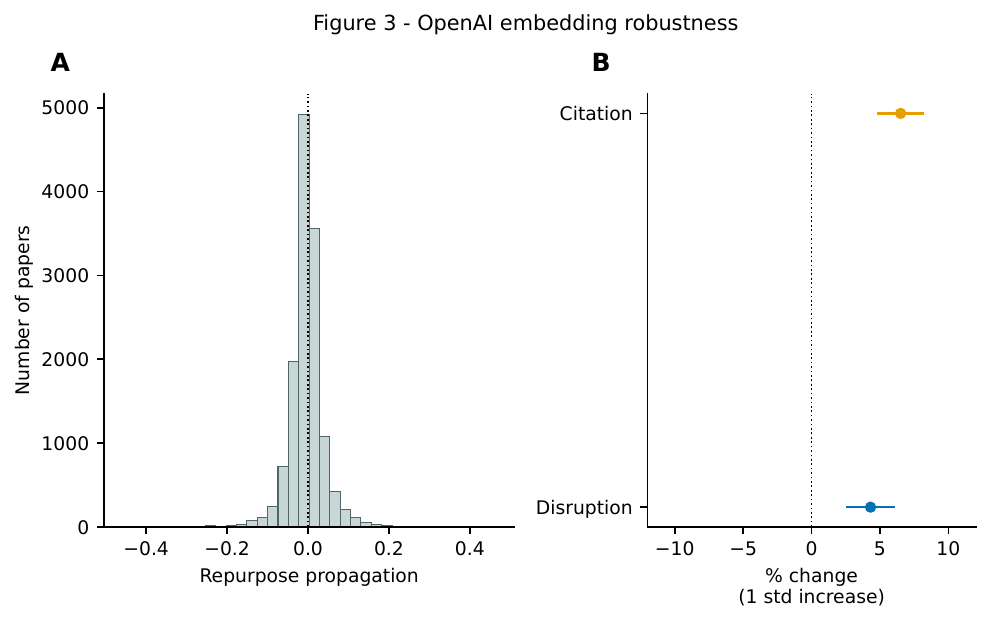}
\caption{
Although repurposed data tend not to propagate, instances of propagation are associated with substantial scientific impact.
(A) Distribution of repurpose propagation. 
(B) Effect size of repurpose propagation on the disruption score over three years, estimated using OLS regression, and on citation count over the same period, estimated using negative binomial regression. Both models include all control variables.
}
\label{fig:a3_openai}
\end{figure}

\begin{figure}[tbhp]
\centering
\includegraphics[width=.9\linewidth]{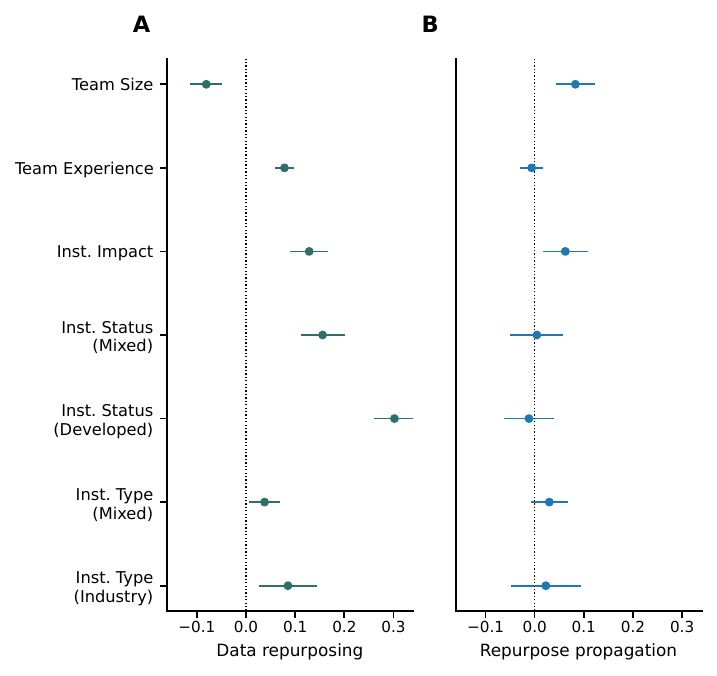}
\caption{
Team characteristics are more strongly associated with data repurposing than with its propagation.
(A) OLS regression results examining the relationships between team size, team experience, institutional impact, institutional global status, and institutional type (industry vs.\ academia) and the degree of data repurposing. 
(B) Relationships between the same team and institutional characteristics and the degree of repurpose propagation. We estimate a separate regression model for each listed variable, controlling for data recombination, the number of datasets used, and dataset usage frequency. Markers represent regression coefficient estimates, and error bars indicate 95\% confidence intervals.
}
\label{fig:a4_openai}
\end{figure}

\newpage
\bibliographystyle{unsrt}
\bibliography{cite_cleaned0711}
 
\end{document}